%% file: main.tex
\documentclass[USenglish,oneside,twocolumn]{article}

\usepackage[listings]{tcolorbox}

\usepackage[utf8]{inputenc}%(only for the pdftex engine)
\usepackage[big]{dgruyter_NEW}
\usepackage{enumitem,kantlipsum}

\usepackage{booktabs}
\usepackage{subcaption}
\usepackage{epsfig,endnotes}
\usepackage{tikz}
\usepackage{filecontents}

\usepackage{graphicx}
\usepackage{lipsum}
\usepackage[linesnumbered,ruled,vlined]{algorithm2e}
\SetKwInOut{Parameter}{Parameters}
\usepackage{amsthm}

\theoremstyle{remark}

\theoremstyle{definition}

\usepackage[absolute]{textpos}
\usepackage{amsmath,amsfonts,amssymb,amsthm}
\usepackage{newtxmath}
\usepackage{mathtools}
\usepackage{commath}
\usepackage{booktabs}
\usepackage{enumitem}
\usepackage{xspace}
\usepackage[vskip=1em,font=itshape,leftmargin=2em,rightmargin=2em]{quoting}

\usepackage[normalem]{ulem} % \sout command
\usepackage{xcolor}
\usepackage{soul}
\usepackage{todonotes}
\usepackage{url}
\usepackage{pifont} % for xmark
\newcommand{\xmark}{\ding{53}}

\newcommand{\app}{VCA}

\renewcommand*{\thefootnote}{\arabic{footnote}}

\newcounter{daggerfootnote}

\newcommand*{\blfootnote}[1]{%
  \begingroup
  \renewcommand\thefootnote{}\footnote{#1}%
  \addtocounter{footnote}{-1}%
  \endgroup
}
\DOI{foobar}

\cclogo{\includegraphics{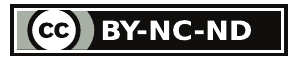}}

\include{include}
 
\begin{document}

% GKT Note: Just use \author*{} if you both want to be corresponding authors.
% Otherwise, it doesn't format nicely.
% I personally recommmend only one corresponding author or to skip using this altogether.

  \author[1]{Yucheng Yang$\dagger$}

  \author[2]{Jack West$\dagger$}

  \author[3]{George K. Thiruvathukal}

  \author[4]{Neil Klingensmith}
  
  \author*[5]{Kassem Fawaz}

  \affil[1]{University of Wisconsin-Madison, E-mail: yang552@wisc.edu}

  \affil[2]{Loyola University Chicago, E-mail: jwest1@luc.edu}

  \affil[3]{Loyola University Chicago, E-mail: gkt@cs.luc.edu}

  \affil[4]{Loyola University Chicago, E-mail: neil@cs.luc.edu}
  
  \affil[5]{University of Wisconsin-Madison, E-mail: kfawaz@wisc.edu}

  \title{\huge Are You Really Muted?: A Privacy Analysis of Mute Buttons in Video Conferencing Apps}

  \runningtitle{Are You Really Muted?: A Privacy Analysis of Mute Buttons in VCAs}

  %\subtitle{...}
\input{PoPETS2022/0_abstract}
%   \keywords{keywords, keywords}
%  \classification[PACS]{}
 % \communicated{...}
 % \dedication{...}

%%===================================================================
\journalname{Proceedings on Privacy Enhancing Technologies}
\DOI{Editor to enter DOI}
\startpage{1}
\received{..}
\revised{..}
\accepted{..}

\journalyear{..}
\journalvolume{..}
\journalissue{..}
%%=================================================================== 

\maketitle
\blfootnote{$\dagger$ Both authors contributed equally to this work.}
\input{PoPETS2022/1_introduction}
\input{PoPETS2022/2_background}
\input{PoPETS2022/3_user_study}

\input{PoPETS2022/4_analysis_mute}
\input{PoPETS2022/4_windows_analysis}

\input{PoPETS2022/5_classification}
\input{PoPETS2022/6_discussion}

\input{PoPETS2022/8_conclusion}

\newpage
\bibliographystyle{plain}
\bibliography{refs,privacyrelated}

\input{PoPETS2022/appendix}

\end{document}

%% file: PoPETS2022/0_abstract.tex
\begin{abstract}
{
Video conferencing apps ({\app}s) make it possible for previously private spaces --- bedrooms, living rooms, and kitchens --- into semi-public extensions of the office.
For the most part, users have accepted these apps in their personal space without much thought about the permission models that govern the use of their private data during meetings.
While access to a device's video camera is carefully controlled, little has been done to ensure the same level of privacy for accessing the microphone.
In this work, we ask the question: \emph{what happens to the microphone data when a user clicks the mute button in a {\app}?}
We first conduct a user study to analyze users' understanding of the permission model of the mute button.
Then, using runtime binary analysis tools, we trace raw audio flow in many popular {\app}s as it traverses the app from the audio driver to the network.
We find fragmented policies for dealing with microphone data among {\app}s --- some continuously monitor the microphone input during mute, and others do so periodically.
One app transmits statistics of the audio to its telemetry servers while the app is muted.
Using network traffic that we intercept en route to the telemetry server, we implement a proof-of-concept background activity classifier and demonstrate the feasibility of inferring the ongoing background activity during a meeting --- cooking, cleaning, typing, etc.
We achieved 81.9\% macro accuracy on identifying six common background activities using intercepted outgoing telemetry packets when a user is muted.
}
\end{abstract}

%% file: PoPETS2022/1_introduction.tex
\vspace{-14mm}
\section{Introduction}

As the de facto alternative for in-person meetings during the COVID-19 pandemic, the demand for online video conferencing for professional and personal use increased significantly. Video Conference Apps (VCAs), such as Zoom, Slack, Teams, and Webex, became available on all modern devices and operating systems. To support their functionality, these VCAs require access to the device's microphone and camera. Operating systems (OSes) provide the users with permission controls that allow the app to access the microphone and camera. Once granted, the app has access to both hardware resources until the user revokes the permission. 

In addition to OS-based controls, VCAs provide their users with two privacy control mechanisms during a call: turning off the camera and muting the microphone. In most OSes, such as Windows and macOS,  turning off the camera from the app engages an OS-level control which prevents the app from accessing the camera. A visible hardware indicator (e.g., a light near the camera) informs the user whether an app is accessing their camera. On the other hand, the implementation of the mute button is app-dependent and rarely has a visible hardware indicator. OSes do not expose an easily accessible microphone switch to the apps without going through many steps (e.g., via a control panel). 

Apart from smart speakers, which pose tangible privacy threats, the mute button has received little attention in the context of VCAs. Previous research investigates users' privacy attitudes towards VCAs and alludes to the mute button as a privacy control tool available to the users during a virtual meeting~\cite{privacyattitudes,6479872}. However, the mute button's privacy implications during the interactions between the user and VCAs have not been adequately addressed.

This paper investigates the privacy issues associated with the mute button in VCAs, focusing on whether a mismatch exists between the user's perception of the mute button and its actual behavior. We follow a two-pronged strategy to guide our investigation.  First, we design a user study to uncover what the users think the mute button does (i.e., their understanding) and what they believe it should do (i.e., their expectations). Second, we compare the user study findings against an empirical investigation of the actual behavior of the mute button across a range of VCAs and operating systems.

We conducted a user study with 223 participants recruited from Prolific. Our user study revealed that the participants perceive the mute button of {\app}s as a privacy control, preventing other meeting participants from overhearing them. We observed a dichotomy in the understanding of the mute button: participants were split about whether a VCA accesses the microphone after they click the mute button. However, most of them indicated that the VCA should access the microphone only when unmuted. 

Based on the findings from the user study, we empirically characterized the conditions in which the VCA \textit{actively} queries the microphone in different operating systems. This task was challenging because OSes only log microphone accesses for each app; they do not provide fine-grained statistics about microphone queries. We addressed this challenge by instrumenting Windows, macOS, Linux, and the Chromium browser to track the fine-grained microphone queries by popular VCAs. We conducted a set of experiments on each VCA-OS combination to monitor the API accesses of each VCA under different conditions. We discovered that \textit{all} of the apps in our study could actively query (i.e., retrieve raw audio) the microphone when the user is muted. Interestingly, in both Windows and macOS, we found that Cisco Webex queries the microphone regardless of the status of the mute button.

We followed our instrumentation efforts with an analysis of Webex, a popular VCA for the enterprise setting. We analyzed how it processes the queried microphone data to determine whether any audio-derived data leaves the device. This analysis also proved challenging as the VCAs, such as Webex, encrypt outgoing traffic. Further, tracking the data flow within apps is not straightforward because they employ proprietary and obfuscated libraries. To facilitate tracking of audio data, we performed a backward search from the encrypted network traffic to locate the inputs to the encryption functions. This search allowed us to decrypt the contents of the network packets sent by Webex to its servers. We discovered that Webex sent periodic packets containing audio-derived telemetry data to its servers, even when the microphone was muted. Although these packets are transmitted at a low rate (once per minute), their audio-derived values correlate with the volume levels of background activities.

To verify our hypothesis, we present a classifier to fingerprint background activities from these telemetry values. Training this classifier was also challenging. Without access to the proprietary algorithm that generates the audio-derived data, it is not feasible to use existing audio datasets to create training data for the classifier. Furthermore, the training data has to represent real-world situations, including realistic noise types and varying volume levels. We address this challenge by collecting Webex-based telemetry data corresponding to more than 200 hours of background activities. Our evaluation of the classifier with over-the-air data shows that telemetry data from Webex can conclusively fingerprint a set of popular user activities, such as music, chatting, and vacuum cleaning. We demonstrate that even with user data that is compressed and transmitted on a minute-by-minute basis, some activities have unique patterns that are discernable in  Webex's telemetry data. 

Our key contributions are as follows.
\begin{itemize}
\item \textit{User Study}: We conduct a user study with 223 VCA participants to assess their understanding and expectations regarding the mute button (Sec.~\ref{sec:user_study}).
   
\item \textit{Audio Access Tracing}: We analyze VCAs' fine-grained access to the microphone; we found that most VCAs have access to audio-derived data even when the user is muted (Sec.~\ref{sec:analysis}).

\item \textit{Webex-based Case Study}: We conduct a thorough system-level study of the Webex Windows client. We discover that, in contradiction to its claims in the privacy policy, Webex sends periodic audio-derived data to its servers (Sec.~\ref{sec:webex_case}).

\item \textit{Background Activity Detection}: We present a design for a machine learning model the infers background activities from Webex's audio-derived data  (Sec.~\ref{sec:bg_classify}).

\item \textit{Mitigation Strategies}: We distill our findings in the form of mitigation strategies that provide users with better control over the mute button (Sec.~\ref{sec:discussion}).
\end{itemize}

% Further, 
% This objective brings out our first technical challenge:\\
% \noindent
% \textbf{Challenge~1:}
% Every app has different data pipelines and proprietary libraries.
% While OSes 
% Extracting raw data from these operating systems requires specialized instrumentation of the audio-related APIs for each operating system.

% , which brings our second technical challenge:\\
% \noindent
% \textbf{Challenge~2:} 
%  making them challenging to analyze. Further, tracking buffers through a proprietary binary, in real-time, and with no access to the source code is not straightforward.
% VCAs employ highly platform-specific methods, which require extensive manual effort to investigate them. In many cases, runtime instrumentation tools slow the execution of the VCAs, causing race conditions that result in crashes, further complicating the tracking task. 

%% file: PoPETS2022/2_background.tex
\section{Related Work}
% \revreq

In the following, we discuss the recent results about the privacy of VCAs. While there is existing research studying possible exfiltration of audio and video data from mobile apps~\cite{PanRenPanoptispy}, we focus on the research specific to VCAs.  We also include related work about mute buttons in the context of smart speakers. Finally, we discuss the work surrounding background activity recognition, which is relevant for our analysis in Sec.~\ref{sec:bg_classify}.

\paragraph*{Privacy Issues in VCAs}

The security and privacy of video conferencing platforms has been studied since the early 2010s. In 2013, Kilpi et al. examined privacy and security issues in future (at that time) videoconferencing technologies~\cite{6479872}. They discuss the mute button as a necessary privacy control for the users. More recently, Emami-Naeini performed an online user study to understand the user concerns with VCAs~\cite{privacyattitudes}. They found that users are concerned about the security and privacy properties of VCAs. They also found that individuals consider the mute button as a privacy control: they perceive privacy violations from forgetting to press the mute button. 

During the pandemic, more people were exposed to privacy and security risks caused by VCAs \cite{zoom_ill,skype_sec}. 
% Camera usage in {\app}s can leak more user privacy but are better protected.
In 2019, Zoom fixed a camera leakage vulnerability caused by its casual use of a local web server~\cite{zoomwebserver}.
Meanwhile, real-time background blurring for {\app}s is widely adopted to protect user's privacy in an office or home environment~\cite{zhang2006light, o2012video}. 
However, {\app}s may leak a user's video privacy in many ways. 
Kagan et al.~\cite{kagan2020zooming} demonstrated that collage images of video conference meetings posted on public websites may leak sensitive information such as users' names, ages and genders.
Altschaffel et al.~\cite{altschaffel2021meta} showed that traffic patterns of encrypted metadata and multimedia data exchanged during {\app} meetings, can be used to identify increased activity in front of camera or even identify users. There are also concerns with the information that VCAs collect about their users. For example, Consumer Reports identified privacy concerns with the data collection practices of popular VCAs, such as Zoom, Google Meet, Microsoft Teams, and Cisco Webex~\cite{consumer_reports}. These concerns centered around the purposes of collecting metadata from the meetings.

In this paper, we follow-up on these previously-reported vulnerabilities and privacy studies. In particular, we study the users' understanding and expectation of the mute button, and whether they match the VCAs' behavior. We focus on the interaction between the VCA and the user's microphone when the user presses the mute button, as opposed to previous research that studies the mute button in the context of protecting the user's privacy from other meeting participants.

\paragraph*{Mute Button in Voice Assistants}
Researchers have also considered the privacy issues from always-listening smart home devices\cite{lau2018alexa,abdi2019more}. Smart home devices continuously process the raw audio to detect a trigger word or phrase. As such, the privacy threats arise from these devices accidentally or maliciously recording the user's background activities~\cite{kws_privacy}. Researchers have first discussed the efficacy of the physical mute button as a privacy control to mitigate these threats. The mute button was found to be inconvenient and suffering from user trust issues~\cite{lau2018alexa, chandrasekaran2018blackout}. Follow-up works proposed other privacy controls, such as ultrasound jamming~\cite{sun2020alexa,chandrasekaran2018blackout,chi_ultrasound}, cutting the power~\cite{chandrasekaran2018blackout}, and employing interpersonal communication cues~\cite{Mhaidli_pets_2020}. 

Contrary to the smart device case, VCA users widely utilize the mute button to prevent others from listening to their background activities (Sec.~\ref{sec:user_study}). Users trust that other meeting participants cannot hear them after applying the mute button. However, the behavior of the VCA, after applying the mute button, is less understood. In this paper, we characterize the operation of the mute button from the perspective of the interaction between the user and the VCA.

\paragraph*{Activity Fingerprinting}
Finally, we discuss research about fingerprinting activities from audio-derived data.
User activities and contextual information, including walking, driving, and riding, can be inferred from ambient sound.
Lu et al.~\cite{lu2009soundsense} presents an audio event classifier that identify user's current activities utilizing the microphone input of mobile phones.
Not only the ambient sound, but encrypted audio traffic can be used to infer user's private information. 
Previous studies proved that encrypted IoT traffic might leak private information of their environment, including device status and user activities. Traffic analysis of the video streams from home security cameras enables monitoring daily activity patterns~\cite{li2020your,apthorpe2017smart,cheng2017homespy}. Li et al.~\cite{li2016side} further demonstrated the possibility of detecting fine-grained activities, including dressing, moving, and eating, from encrypted home security camera traffic. They selected features such as traffic packet size and length distribution. 
Similar to encrypted traffic analysis, Schuster et al.\cite{schuster2017beauty} performed an encrypted Video Stream Identification by analyzing bitrate burst and time interval of video streaming traffic. They utilized the segment transmission mechanism of MPEG-DASH and successfully identified Netflix video titles using a trained classifier.

Kennedy et al. and Wang et al.~\cite{kennedy2019can, wang2020fingerprinting} demonstrated that an attacker can infer which voice commands a user says to a smart speaker, by eavesdropping and analyzing outgoing encrypted traffic from smart speakers to a cloud server. 
Wang et al.~\cite{wang2020fingerprinting} further manifested the incoming traffic from the server also leak voice commands information. 
Moreover, Bae et al.~\cite{277230} presented a video streaming service identification attack by monitoring video downstreaming traffic through LTE networks with high accuracy.

These research works demonstrate that data derived from audio streams can be used to fingerprint their content and is therefore relevant to our discussion in Sec.~\ref{sec:bg_classify} about inferring the background activities while the user is muted.

% Meanwhile, IoT traffic between devices can leak user activities and device status as well.  Through sniffing wireless traffic between IoT devices and device to gateways, Acar et al. \cite{acar2020peek,pinheiro2019identifying,gu2020iotspy} demonstrate that attackers can infer IoT device types, the on/off status of devices, and even the ongoing user activities.

% Related work: ignored audio privacy in PC devices in a smart home environment, the microphone may leak audio data when the app is running, during a meeting, during a meeting muted

% A study done by Pan et al.~\cite{PanRenPanoptispy}, revealed that third-party applications running in Android OS showed signs of suspicious behavior, implying they may be collecting user data.
% The study did not directly address privacy problems in {\app}s, however the applications that they deemed suspicious shared the same access permission model given to {\app}s.

%% file: PoPETS2022/3_user_study.tex
\input{PoPETS2022/3_surveyquestions}

\section{User Study}
\label{sec:user_study}

Our first objective is to study the user perceptions of the mute button along with their understanding of its functionality. Towards that end, we conduct an online user study with 230 VCA users. Our study aims to answer two questions about VCA users: (1) \textit{When do they think the VCA accesses their microphone?} and (2) \textit{When do they think the VCA should access their microphone?}. Answering these questions allows us to characterize the user’s understanding and expectations of the mute button, respectively. In the following, we describe the design of the user study, the recruitment, and the findings.

\subsection{Study Design}

We designed a Qualtrics survey\footnote{The full survey can be found here: \url{https://osf.io/szd4x/}} to help answer our research questions. We used partial disclosure to hide the fact that the study was about the privacy implications of the mute button.
The description of the survey and its title focus on capturing the users' general experience with VCAs during the pandemic. The survey has four major sections; the first section collects optional demographic information. The second section collects information about the preferred VCA and frequency of usage. 

The third section asks the respondents about their experience with the mute button. We adapt the questions from Lau et al.~\cite{lau2018alexa}, which studies the mute button in smart speakers. In particular, we probe the users about their usage of the mute button, their reasons, and their understanding of its functionality using questions in Table~\ref{tab:survey_qs}. This section contains three open-ended questions and two multiple-choice questions.

The last section adopts a refined version of Internet Users’ Information Privacy Concerns (IUIPC-8) from Gro{\ss}~\cite{IUPIC} to measure the participants' privacy concern. This survey section contains the first mention of privacy, after the respondents have answered the questions related to the mute button. Finally, the survey includes two attention checker questions and was exempted by the IRB at our institution.

\paragraph*{Participant Recruitment and Demographics}
We recruited participants from the Prolific data collection platform. We employed Prolific's prescreening criteria to enforce gender balance and to forward the survey to only those who have worked from home during the COVID-19 pandemic with 90\% approval rate in previous studies. Before conducting the survey, we conducted a pilot study with 15 users to calibrate the payment and ensure that the study design is clear. Through Prolific, we were able to recruit 299 participants, where we kept 223 responses from participants who passed the attention checkers. The median completion time was 8 minutes, and we paid each participant \$1.5; the median hourly rate was \$11.

% \todo[inline]{@Yucheng: we need to update the numbers below}
% \todo[inline]{@Yucheng, can you make a demographics chart to accompany the explanation?}
Among our participants, 96.8\% are between 18 to 44 years old, 63.2\% of them work in sales, service, management and professional industry, and 82.5\% achieved at least a college degree. During COVID-19, 54.7\% of our participants answered that they have used video conferencing apps more than once a day and 40\% of them used once a day or once every few days. The most popular video-conferencing app among the participants is Zoom, and the other popular apps include Microsoft Teams, Google Meet and Cisco Webex.

We map the responses to the IUIPC-8 question to a score based on seven-point Likert scale, representing participant's privacy attitudes. The average scores is 2.02 for all participants, implying that most participants are privacy-conscious in our study. The value of Cronbach Alpha Index is 0.7915 for privacy attitudes responses from 223 participants, which indicates a good internal consistency and reliability of these responses.

\subsection{Findings}
We report the key findings from our user study, through analyzing the participants' responses. We coded the responses to the open-ended questions (\textit{Q1}, \textit{Q2}, and \textit{Q3}) following this procedure. For each question, two authors independently coded the responses, after which they generated a consolidated codebook describing the responses. For \textit{Q1}, we settled on five codes about the reasons for which participants use the mute button. For \textit{Q2}, the codebook consists of twelve codes describing the background activities. The codebook for \textit{Q3} contains nine codes representing the participants' description of the mute button operation. Then, each coder independently coded the first 30 responses for each question; the resulting Cohen's kappa is 0.85 for \textit{Q1}, 0.90 for \textit{Q2}, and 0.82 for \textit{Q3}, indicating strong agreement~\cite{pmid23092060}. The coders split and coded the rest of the responses. See detailed codebooks of the open-ended questions in Appendix~\ref{sec:codebook}.\\

\begin{figure}[t]
    % \centering
    \includegraphics[width=\linewidth]{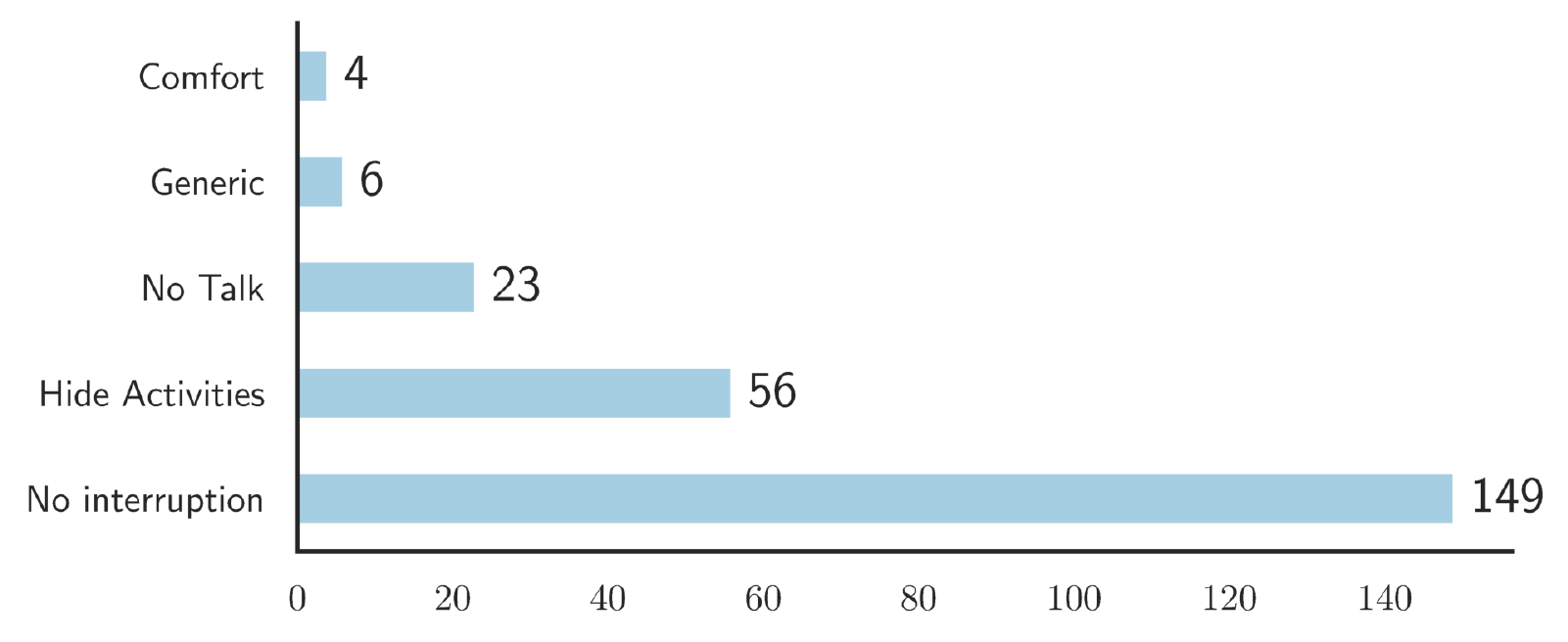}
    \caption{The distribution of the codes about reasons users reported for using the mute button as extracted from answers to \textit{Q1}.}
    \label{fig:codes_q1}
\end{figure}

\begin{figure}[t]
    % \centering
    \includegraphics[width=\linewidth]{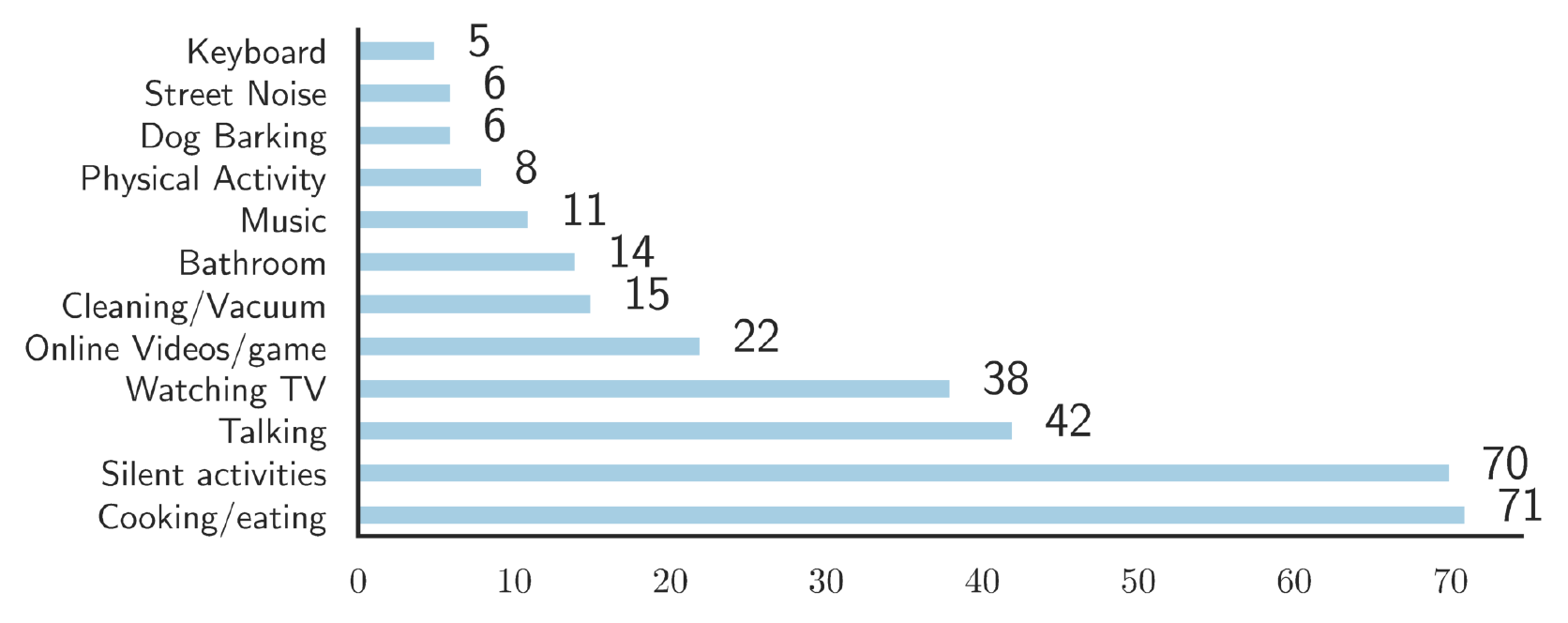}
    \caption{The distribution of the codes about the background activities as extracted from answers to \textit{Q2}.}
    \label{fig:codes_q2}
\end{figure}

\noindent
\textbf{Usage Patterns:}
We start by analyzing the responses to \textit{Q1}, where 214 participants out of 223 indicated that they have used the mute button before. The responses for \textit{Q1}, as shown in Fig. \ref{fig:codes_q1}, reveal two main reasons why users employ the mute button: (1) hide background activities and (2) avoid interrupting or disturbing others on the call. It is interesting that the participants regard the mute button as a privacy control measure to prevent others from hearing them. For example, \textit{P19} mentioned the reason for using the mute button is: \textit{``So that people won't listen to private activities or conversations.''} 

The responses for \textit{Q2} indicate an array of background activities the participants perform while muted, as indicated in Fig.~\ref{fig:codes_q2}. Participants mentioned more than one activity in their responses; For example, \textit{P166} mentioned: \textit{``Talking, loud video watching, cat activity (meows, occasional falling and crashing of items), cleaning (including vacuuming).''} The most prevalent activity was related to preparing food, cooking, snacking, or eating. Other frequent activities include chatting, watching TV, cleaning, typing, or watching online videos. We elaborate more on these background activities in Sec.~\ref{sec:bg_classify}. \\

\begin{figure}[t]
    % \centering
    \includegraphics[width=\linewidth]{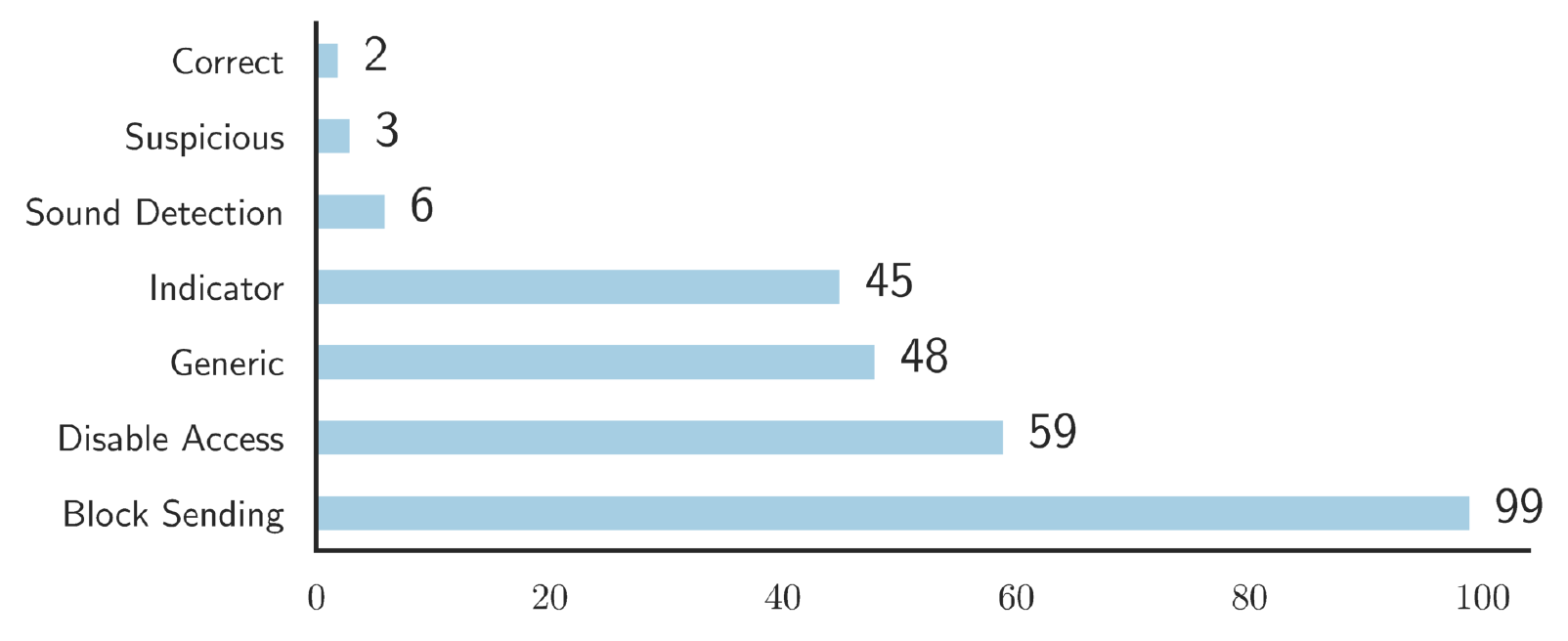}
    \caption{The distribution of the codes about the users' understanding of the mute button operation from answers to \textit{Q3}.}
    \label{fig:codes_q3}
\end{figure}

\noindent
\textbf{Understanding of the Mute Button:}
As indicated earlier, we asked the participants two questions (\textit{Q3} and \textit{Q4}) to gauge their understanding of the mute button. To gain initial insights into the participants' understanding of the mute button, we study the coded responses to \textit{Q3}, as evident from Fig.~\ref{fig:codes_q3}. The most frequent response was that by using the mute button, the app prevents others in the call from hearing the user. For example, \textit{P16} indicated that: \textit{``It doesn't produce my audio on the other participant's platform or computer.''} Moreover, other participants focused on the interface change when the mute button is pressed, as in the case of \textit{P119}: \textit{``It shows me the mic with a line crossing it signalling it is not working.''} . Meanwhile, 59 participants mention that the mute button disables the microphone. For example, P161 mentions: \textit{``When I press the mute button, my microphone is muted and disabled on the app from picking up any sound waves from where I am.''}
% We found that this interface change gives the participants the impression that the microphone access is disabled or cut to the app

\begin{figure}[t]
    % \centering
    \includegraphics[width=\linewidth]{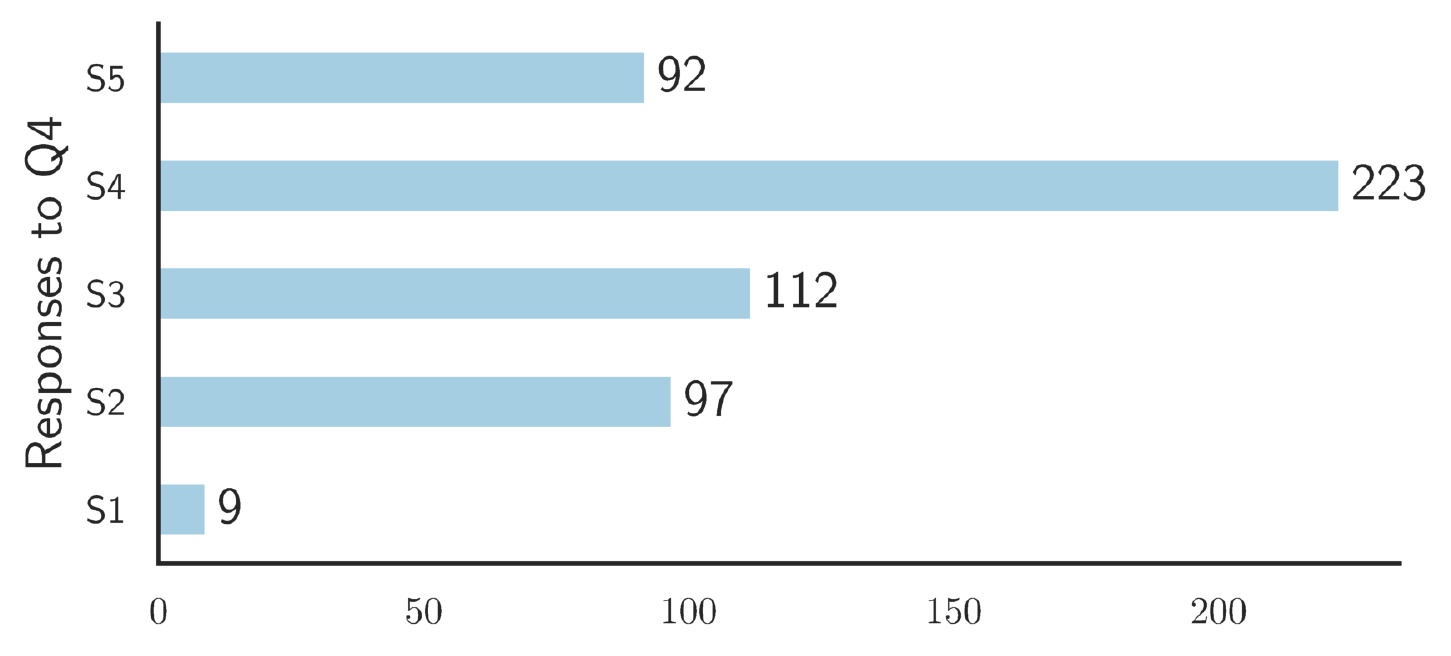}
    \caption{The distribution of responses to Q4. The statements \textit{S1-S5} are defined in Table~\ref{tab:survey_qs}.} 
    \label{fig:pa}
\end{figure}

For Q4, we provide five situations, \textit{S1-S5}, in our user study as shown in Table~\ref{tab:survey_qs}.
The responses to Q4 indicate that the participants exhibit a diverse understanding of the operation of the mute button, as shown in Fig. \ref{fig:pa}. Out of the 223 responses, 69 participants selected only \textit{S4} as a response to Q4. These participants think the app only accesses the microphone when they are in the meeting and the mute button is not pressed. 

Further, we found that the participants were split in their selection of \textit{S3} as a response to Q4. Nearly half of the participants (111) did not select \textit{S3}, indicating that the app does not access the microphone when the mute button is pressed. The other half indicated that the app accesses the microphone, even when muted. Interestingly, we observe that 49 participants selected \textit{S2}, \textit{S3}, \textit{S4}, \textit{S5} when responding to Q4, indicating that the app accesses the microphone as long as it is running. Also, we observe that 36 participants selected \textit{S3} and \textit{S4},  indicating that the app accesses the microphone as long as the user is in a meeting. In all the cases above, we found no correlation between the responses and the IUIPC-8 privacy attitude scores. \\

\begin{figure}[t]
    % \centering
    \includegraphics[width=\linewidth]{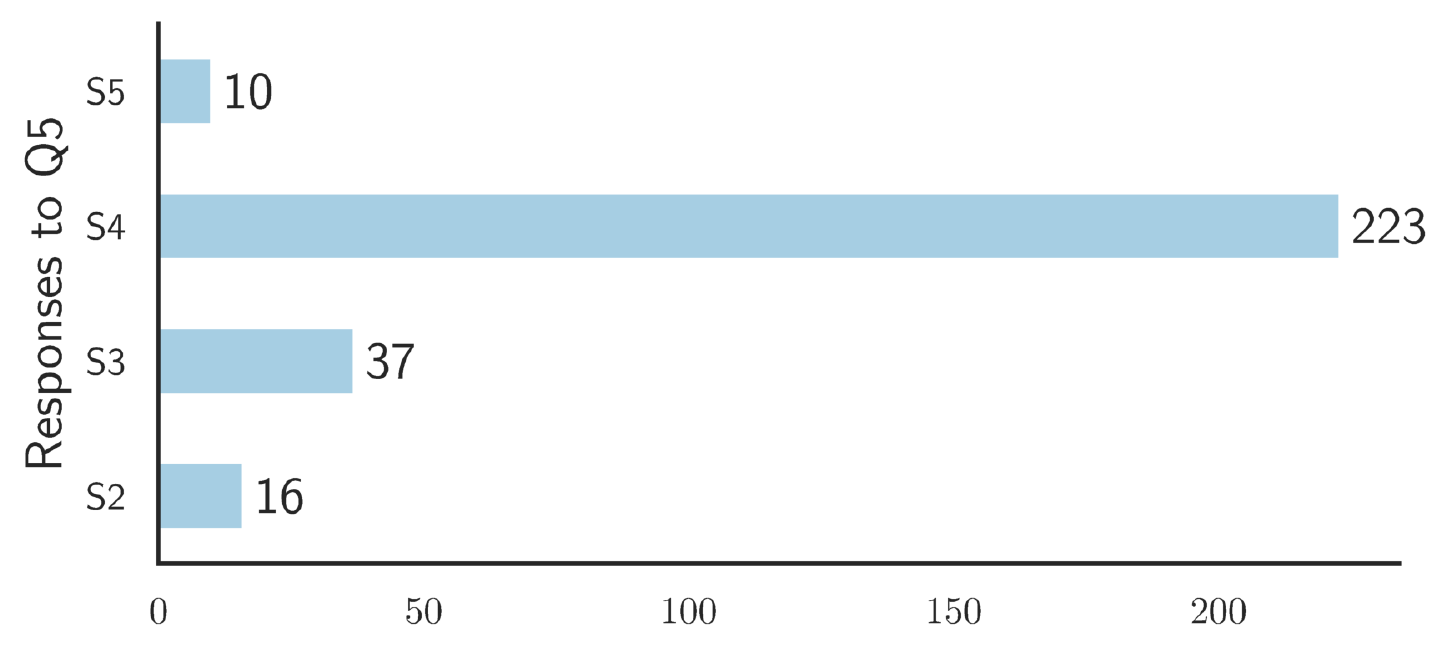}
    \caption{The distribution of responses to Q5. The statements \textit{S1-S5} are defined in Table~\ref{tab:survey_qs}.} 
    \label{fig:q5}
\end{figure}

\noindent
\textbf{Expectations of the Mute Button:}
Finally, we analyze the responses to \textit{Q5}, about when do the participants think the VCAs should access the microphone. The responses reveal that the participants have clear expectations about the operation of the mute button, as indicated in Fig. \ref{fig:q5}. Among the 223 responses, 173 participants selected only \textit{S4} as a response to \textit{Q5}. These participants indicated that the app should only access the microphone when the meeting is running and the user is unmuted. Interestingly, 27 respondents selected both \textit{S3} and \textit{S4} as a response to \textit{Q5}. 

In conclusion, the results from the user study suggest that the user's understanding of the mute button does not match their expectations of its behavior. In the rest of this paper, we study the actual behavior of the mute button and analyze whether it matches user understanding and expectations.

%% file: PoPETS2022/3_surveyquestions.tex
 \begin{table*}[t]
    \begin{subtable}{0.29\textwidth}
        \centering
        \begin{tabular}{p{4.8cm}}
        \multicolumn{1}{c}{\textbf{Open Ended Questions}}\\ \midrule
        \textbf{Q1.}  Why do you use the mute button? \\
        \textbf{Q2.}  What activities do you perform or take place in your background when you are muted? \\
        \textbf{Q3.} Please describe what does the app do when you press the mute button.    \\
        \end{tabular}
    %   \caption{First Week}
       \label{tab:open_ended}
    \end{subtable}
     \hfill
    \begin{subtable}{0.28\textwidth}
        \centering
        \begin{tabular}{p{4.6cm}}
        \multicolumn{1}{c}{\textbf{Multiple Choice Questions}} \\ \midrule
        \textbf{Q4.} For your most frequently used video meeting app, when do you think it has access to your microphone? \\
       \textbf{Q5.} For your most frequently used video meeting app, when should it have access to your microphone?\\
        \end{tabular}
        % \caption{Second Week}
        \label{tab:multiple}
     \end{subtable}
      \hfill
    \begin{subtable}{0.4\textwidth}
        \centering
        \begin{tabular}{p{7cm}}
        \multicolumn{1}{c}{\textbf{Multiple Choice Answers for Q4 and Q5}} \\ \midrule
        \textbf{S1.} When the app is not running. \\
        \textbf{S2.} You start the app but are not in a meeting. \\
        \textbf{S3.} You're in a meeting but you apply the mute button in the app. \\
        \textbf{S4.} You're in a meeting and you are unmuted. \\
        \textbf{S5.} You leave the meeting while the app is still running.\\
        \end{tabular}
        % \caption{Second Week}
        \label{tab:answers_multiple}
     \end{subtable}
     \caption{The main questions used in the user study. Q1-Q3 are open-ended questions with answers coded by researchers. Q4 and Q5 are multiple choice questions where the participant selects one or more statements from \textit{S1-S5} in response. The full list of questions is available in the Appendix.}
     \label{tab:survey_qs}
\end{table*}

%% file: PoPETS2022/4_analysis_mute.tex
\section{Analysis of Mute Button}
\label{sec:analysis}
% \todo[inline]{File a disclosure report to Webex!}
% \reved

Following the results from our user study, we investigate whether the actual behavior of {\app}s matches user expectation by focusing on desktop environments.  
Our objectives are to determine: (1) if VCAs actively access the microphone when muted and (2) what kind of indicators (if any) they give users that the microphone is being accessed.

\subsection{Overview of VCAs and Platforms}

There are two broad categories of runtime environments in which VCAs execute: native apps that run directly in the operating system and web apps hosted by a web browser.
Each has a different permission model for accessing the microphone.
Most of the VCAs we study in this work have a native app implementation for the major operating systems (macOS and Windows) and a web app used on unsupported platforms (Linux and others). The VCAs that we studied (listed in Table~\ref{tab:implementations}) exhibit a consistent look and feel across platforms. Their implementation, however, on each platform is different, due to syscall interfaces and display APIs. Zoom on Windows, for example, is a self-contained Windows-specific software package. Zoom on macOS has a similar user interface to its Windows counterpart, but the underlying code base appears to be different. 

Native apps can collect data from the microphone with few restrictions. Web apps---implemented in JavaScript--- request access to the microphone through a web browser, which generally has more restrictive policies for data collection and more tools that allow the user to control the app's access to hardware.

% The differing codebases are due to the implementation details in the operating system, such as syscall interfaces and display APIs.

\begin{table}[t]
\centering
\begin{tabular}{lcccc}
% \toprule
\multicolumn{1}{c}{App} & Windows     & Linux      & macOS        \\ \midrule
Zoom (Enterprise)       & \checkmark  & \checkmark & \checkmark    \\
Slack                   & \checkmark  & \checkmark & \checkmark   \\
MS Teams/Skype   & \checkmark  & \xmark     & \checkmark          \\
Google Meet             & $\circ$     & $\circ$    & $\circ$         \\
Cisco Webex             & \checkmark  & $\circ$    & \checkmark     \\
BlueJeans               & \checkmark  & $\circ$    & \checkmark     \\
WhereBy                 & $\circ$     & $\circ$    & \checkmark    \\
GoToMeeting             & \checkmark  & $\circ$    & \checkmark     \\
Jitsi Meet              & \checkmark  & $\circ$    & \checkmark     \\
Discord                 & $\circ$     &\checkmark  & $\circ$         \\ 
% \bottomrule
\end{tabular}
\caption{A summary of the VCAs we studied. \checkmark: native app\\ $\circ$: web-based app \xmark: No implementation. }
\label{tab:implementations}
\end{table}

\paragraph*{Browser Based Apps}

Browser-based VCAs rely on their host browser to mediate their interactions with the operating system and the hardware.
The browser-based VCAs that we studied are implemented entirely in JavaScript, and they use a special-purpose API called WebRTC~\cite{webrtc} for driver interactions---including microphone accesses---that are typically not available to web apps.
WebRTC is a native interface written in C++ and C, acting as a driver for the hardware within the browser that can call the operating system to access the microphone.
Information transferred by WebRTC is subject to controls and policies of the browser. Web-based VCAs are sandboxed inside the browser and do not circumvent WebRTC.

There are two ways a user can mute a web-based VCA: (1) using a browser-level mute button or (2) using a WebRTC software mute signal from the app.
Both techniques are more trustworthy than app-controlled mute because they are implemented and enforced by the browser, not the app.

The browser-level mute button completely disables microphone access to the VCA, as if the microphone is not active within the system.
Web-based VCAs also implement an app-level mute button, which has similar functionality to the browser-level mute: it enables a software mute inside of WebRTC, disabling all audio transfers from the microphone.
Users must trust the web-based VCA to use the software mute functionality rather than some internal mute button implementation. We found that all of the studied apps use the WebRTC mute functionality correctly.
Furthermore, it is straightforward to verify that web-based VCAs correctly use the software mute functionality through source code audits and the WebRTC debugger built into Chromium.

\paragraph*{Native Video Conferencing Apps}

Native VCAs can directly call the operating system to retrieve audio data from the microphone.
Most of them abide by the operating system (OS) rules to access the microphone data, with some exceptions.
The OS imposes fewer restrictions on native apps than the browser runtime environment imposes on web apps.

All operating systems utilize a permissions-based access system to retrieve data from the microphone.
In most cases, apps must have explicit permission to access hardware resources such as the microphone. Each app follows three steps to configure and use the microphone: (1) user approval, (2) driver initialization, and (3) audio data retrieval.
Windows and macOS require the user to explicitly provide permission for each app, which the app retains indefinitely while it runs (unless the user revokes the permission).

Once the user approves the access for the app, the app must create an interface to the audio drivers.
Some OSes, like Windows, offer users a visual cue that indicates when the app is using the microphone.
But unlike the WebRTC browser runtime, none of the major operating systems we are aware of support  enforce a software mute.
This lack of an OS-mediated software mute means each native app must implement its own internal mute functionality.
% The audio retrieval process is not as closely monitored as it is in the browser-based setting.
Even when a software mute is active, apps can still access the microphone while the user is muted.

\subsection{Analysis Methodology}
\label{sec:methods}

To understand what happens when the user presses the mute button on desktop VCA clients, we utilize various OS-based tools to trace audio data as it is transferred from the operating system to the app.
Our objective is not just to establish whether the app has permission to access the microphone when muted.
Instead, we aim to understand whether the app actually reads microphone data when the user is muted.

\paragraph*{Linux}
Audio data transfer from the Linux kernel to the {\app}s is mediated through PulseAudio and ALSA.
ALSA is a kernel subsystem that provides a kernel-level interface to the audio hardware, and PulseAudio is a userland process that interfaces with ALSA and provides higher-level features like mixing and multiplexing.
All the {\app}s we studied interface with the userland PulseAudio process.

To intercept audio data in transit from PulseAudio to a {\app}, we use the DynamoRIO runtime code manipulation system~\cite{dynamorio}, which allows us to inject foreign code into a running process.
Our additional code, written in C, is called each time a fresh buffer of microphone data arrives from PulseAudio.
We write the audio buffer's address in the process's memory space to a log file.
We then trace the buffer addresses from the log using IDA Pro.
The contents of the buffer are the raw audio bytes from the microphone.
DynamoRIO oversees the process's execution by loading and running modified basic blocks one at a time, which substantially slows the app's execution, occasionally causing it to crash.

%\todo[inline]{@Jack: is DynamoRIO slow?  -- Yes ok thx is it shitty in any other way? It is not easy to use There is little documentation and the guy who made it designed it for widnows so the linux build kinda ehh gotcha. does the injected code have to be written in assembly? No just C. Where do the print statements print out to? terminal? You can print it to the terminal or write to a file. Their print statements go either way. nice can you provide a sample of foreign code ?   ya sure let me get you a github link nice thanks https://github.com/DynamoRIO/dynamorio/blob/master/api/samples/memtrace_simple.c or this one https://github.com/DynamoRIO/dynamorio/blob/master/core/win32/inject.c

%wow this code is long i thought it would be like 2 lines I have some that is shorter can i have that? Its on my desktop oh ok sweet. maybe we can put it in a figure in the paper? or  in Ya for sure. I'll send my file over tonight when I get home. I remeber where all of that code is ok cool}

%Once a piece of software establishes a line with the microphone device, the raw audio bytes can be directly copied from a live stream of bytes.
% To capture when a device copies audio data, we have to perform a Systrace on the process. 
% Systrace(strace) allows a user to capture system interrupts from a specified application.
% We used strace to examine if an application is reading data from the microphone while the user was muted.

%, making it difficult to trace calls to the operating system.
%applications query an API to preform specific calls to low level utilities.

\paragraph*{Windows}

Although it is possible to track microphone access by monitoring the system registry~\cite{registrymic}, we were not able to track transfers in real time from the microphone to the {\app}.
The registry only records times at which an app opens or closes a connection to an audio device.
The OS registry---linked to a visual indicator in the system tray---does not distinguish detailed API calls which encode information about whether a {\app} is reading audio data or accessing status flags about microphone activity.% a user details about their audio bytes from the microphone to the app in real-time, nor does it record the raw data from the microphone.
For fine-grained and detailed information, we intercept syscalls from the {\app} to the operating system.

In Windows 10, syscalls are obfuscated behind a userland API library which acts as an intermediary between the apps and the OS.
The Windows API library is similar to the Linux/Unix C library syscall wrappers, except that there is no one-to-one mapping between the parameters that the app passes to the API and the parameters that the API passes to the OS.
Instead, the API functions as a higher-level wrapper around system calls, and there is no official documentation available from Microsoft detailing how to call the operating system directly.

Windows implements many special-purpose API functions for actions like accessing the microphone, which in Linux and Unix are all handled as files.
We develop a two-step process to trace audio data in transit from the Windows OS to the native VCAs.
First, we use a tool called API Monitor~\cite{apimonitor} to instrument the userland API with hooks to log pointers to the inputs and outputs of several microphone-related API calls. We then use a live binary analysis tool called \texttt{x64dbg}~\cite{x64dbg} to read the contents of the buffers out to a log file. We utilize an anti-anti-debugging library called Scylla-Hide~\cite{scyllahide}, which hides the fact that an app is being debugged to prevent the app from crashing.

\paragraph*{Chromium}

Chromium acts as an intermediate layer between the operating system and the browser based VCAs.
To verify whether web-based VCAs access the microphone while muted, we inject our own logging code in the source of Chromium.
We instrument the following three browser functions in Chromium, which are responsible for transporting audio from the operating system to the VCA\footnote{Appendix~\ref{sec:apiappendix} includes more details about the functions inside Chromium.}.
First, the browser initiates audio-related \texttt{read\_data} function, which retrieves the raw microphone data from the operating system and stores it in a raw audio buffer.
Then it calls \texttt{encode} and \texttt{send\_stream} functions, which transforms the raw audio into an encoded stream and transfers the encoded audio stream to the web-based VCAs.

% \begin{enumerate}[topsep=5pt]
%     \item \textbf{Read Data}: retrieves the raw microphone data from the operating system.
%     It is also contains the raw audio buffer. 
%     \item \textbf{Encode}: transforms the raw audio into an encoded stream.  
%     \item \textbf{Send Stream}: Transfers the encoded audio stream to the web-based VCA.
% \end{enumerate}

% \paragraph*{Android}

% We utilize Android's \texttt{dumpsys} tool~\cite{dumpsys} to retrieve diagnostic data for the system's audio service.
% The \texttt{dumpsys} tool produces log files that record each time a VCA calls \texttt{requestAudioFocus()} and \texttt{abandonAudioFocus()}, the functions that Android apps use to capture input from the microphone device in Android 8.0 through Android 11.
% Parameters that the apps pass to \texttt{requestAudioFocus()} or \texttt{abandonAudioFocus()} tell the operating system whether the app is activating and requesting access to the microphone or deactivating the microphone. 

\paragraph*{macOS}

% Audio data transfer from the Linux kernel to the {\app}s are mediated by PulseAudio and ALSA.
% ALSA is a kernel subsystem that provides a kernel-level interface to the audio hardware, and PulseAudio is a userland process that interfaces with ALSA and provides higher level features like mixing and multiplexing.
% The {\app}s we studied all interface with the userland PulseAudio process.

% To intercept audio data in transit from PulseAudio to a {\app}, we used the DynamoRIO runtime code manipulation system~\cite{dynamorio}, which allows us to inject foreign code into a running process.
% Our additional code, written in C, is called each time a fresh buffer of microphone data arrives from PulseAudio.
% We write the audio buffer's address in the process's memory space to a log file.
% We then trace the buffer addresses from the log using IDA Pro.
% The contents of the buffer are the raw audio from the microphone.
% DynamoRIO oversees the process's execution by loading and running modified basic blocks one at a time, which substantially slows the app's execution, occasionally causing it to crash.

An audio subsystem manages microphone data created by Apple via \texttt{AVFAudio} or the \texttt{AVAudioEngine} interfaces~\cite{apple_recording}.
These interfaces have the same purpose and interact with the audio hardware in userland.
{\app}s make a system call to \verb"mach_msg_trap" within either an audio interface thread managed by Apple and retrieve raw audio bytes from the microphone.
All of the {\app}s we studied connect to the microphone using either of these interfaces and make the same system calls when reading bytes from the microphone.

To monitor {\app}s' microphone accesses we use a XCode tool called \texttt{Instruments}~\cite{instruments} and the standard Unix networking tool \texttt{tcpdump}.
\texttt{Instruments} logs all system calls and their arguments to a user interface in the Apple system log.
\texttt{tcpdump} records network traffic while any of the {\app}s are running.
We attach \texttt{Instruments} to a live {\app} and perform a \texttt{tcpdump} on the networking interface to extract and monitor the dataflow from microphone to the {\app}.
We then observe the results from \texttt{Instruments} to correlate behavior patterns with with Windows evaluation.
{\app}s in macOS behave similarly to their Windows implementations.

\subsection{Findings}

To understand how VCAs consume microphone data, we conducted experiments on each app-OS combo shown in Table~\ref{tab:implementations}.
We installed all {\app}s and registered two accounts for each app on each of the four operating systems.
The app-OS combinations that are only accessible in a browser are tested in Linux on Chromium.
We initiated the meeting app for each meeting experiment and used the techniques explained above to trace microphone data from OS to VCA under two conditions: mute button toggled on and mute button toggled off. 
Most platforms we studied display a visual indicator to alert the user that an app is accessing the microphone\footnote{Some Linux distributions do not provide any visual indication that the microphone is in use.}.
We found three broad policies that VCAs follow to read data from the microphone while muted:

% in current iOS, imminent Android 12, in macOS12 Monterey
% Apple previously introduced one orange light indicator for microphone usage and recently announced a software-based microphone indicator in macOS 12 Monterey, while Android announced microphone access indicator in Android 12 recently.

%\todo[inline]{clean the following to remove android. add webex for continuously sampling mic. }
\begin{enumerate}
%% webex here
    \item \textbf{Continuously sampling audio from the microphone:} apps stream data from the microphone in the same way as they would if they were not muted.
    Webex is the only {\app} that continuously samples the microphone while the user is muted.
    In this mode, the microphone status indicator from an operating system remains continuously illuminated.

    \item \textbf{Audio data stream is accessible but not accessed:} apps have permissions to sample the microphone and read data; but instead of reading raw bytes they only check the microphone's status flags: \emph{silent}, \emph{data discontinuity}, and \emph{timestamp error}.
    We assume that the VCAs, like Zoom, are primarily interested in the \emph{silent} flag to tell if a user is talking while the software mute is active. 
    In this mode, apps do not read a continuous real-time stream of data in the same way as they would while unmuted.
    Most Windows and macOS native apps\footnote{Except Skype and Teams, which we cannot observe because they do not use the conventional Windows API.} can check if a users is talking even while muted but do not continuously sample audio in the same way as they would while unmuted.
    In this mode, the microphone status indicator in Windows and macOS remains continuously illuminated, reporting that the app has access to the microphone.
    We found that applications in this state do not show any evidence of raw audio data being accessed through the API.
    
    \item \textbf{Software mute:} apps instruct the microphone driver to completely cut off microphone data.
    All of the web-based apps we studied used the browser's software mute feature.
    In this mode, the microphone status indicator in the browser goes away when the app is muted, indicating that the app is not accessing the microphone.
\end{enumerate}

The notable exceptions to these trends are the Microsoft VCAs (Teams and Skype) and Cisco Webex.
Microsoft VCAs are much more difficult to trace because they do not use the standard Windows userland API.
Instead, they directly make calls to the operating system.
Since the Windows syscall interface is undocumented, we could not determine how Teams and Skype use microphone data when muted. More interestingly, we observe that Cisco Webex --- unlike the rest of the Windows native VCAs --- continuously accesses the microphone while muted. Using \texttt{x64dbg}, we were able to trace Webex's copied audio buffer until that buffer reaches the stack. We discovered that while the app was muted, Webex's audio buffer contains raw audio from the microphone. In the next section, we focus our data flow analysis on Cisco Webex in Windows because of its popularity\footnote{The monthly statistics from Cisco Webex include 600 million participants and 6 billion calls~\cite{cisco_usage}.} in the enterprise setting and, more importantly, its unusual behavior.

Recall that our user study reveals two main observations: participants are split whether the VCAs access their microphone while muted, and expect them to access the microphone only when they are unmuted. Our results from this section indicate that the participants are largely unaware of the operation of the VCAs. More importantly, the behavior of these apps violates user expectations. This mismatch between user expectations and app behavior highlights privacy issues with the design of the mute button.

%% file: PoPETS2022/4_windows_analysis.tex
\section{Webex Case Study}
\label{sec:webex_case}

\begin{figure*}[t]
    \centering
    \includegraphics[width=\linewidth]{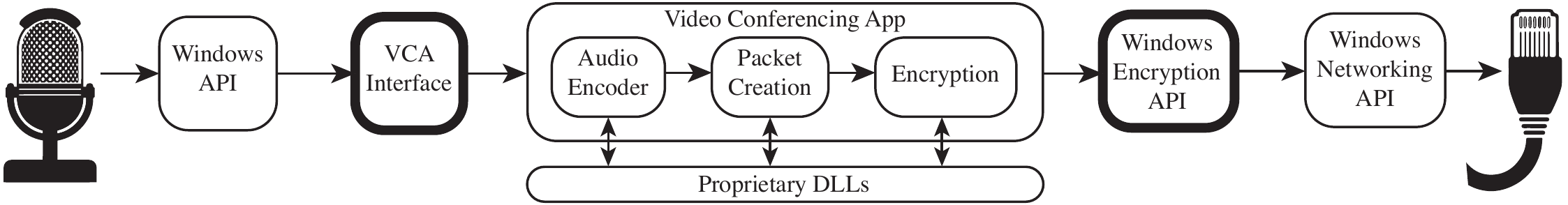}
    \caption{Data flow of audio bytes within a Windows 10 VCA. This pipeline is generalizable across the Windows platform. Our system attaches to the bolded modules.}
    \label{fig:Data_Flow}
\end{figure*}

Based on our findings from the previous section, we perform an in-depth analysis of the microphone access pattern in Cisco Webex\footnote{We used Webex client version 41.12.3.11 through our study.}. We focus on Windows 10 as it is the most widely used operating system at home and in enterprise\footnote{90\% of respondents to our user study used Windows.}. As Webex continuously samples the user's microphone (when muted), we need to study whether audio-derived data leaves the local device.

Determining whether audio-derived data from a \app~ is leaving on the network port is not a straightforward task because the network flow from VCA to a server is encrypted.
Raw dumps of the network traffic from Wireshark are not informative about precisely what the network traffic carries to the {\app}'s server.
And we know that {\app}s send and receive network packets that do not contain any audio or video data, so counting network packets cannot give us an indication of whether audio-derived data is leaving a device while the {\app} is muted.
Instead of directly logging network packets, we need to track how audio data is processed within a \app .

%%this reads a bit informal, needs some tightening. 
\subsection{Methodology for Traffic Interception}

% As we discussed in \S~\ref{sec:analysis}, native Windows apps call a userland API to invoke OS functions like encryption and network transmission.

% By instrumenting the API with hooks, we can force it to leak app data like the locations of specific buffers, which we used to track the microphone data.
% But the chain of operations that {\app}s use to transmit over the network is much longer and involves more software modules than the microphone interface.

% To 
Fig.~\ref{fig:Data_Flow} depicts the flow of data from microphone to network in native Windows apps. 
Understanding how a particular {\app} handles data from the microphone requires tracking the data as it traverses the chain of processing shown in Fig.~\ref{fig:Data_Flow}. 
Most of the data processing in the {\app}s we study is handled by proprietary DLLs\footnote{Dynamically Linked Libraries (DLLs) are the Windows implementation of shared libraries.}.
Tracking data through function calls from the main {\app} process to a DLL is unreliable because runtime binary analysis tools like IDA Pro~\cite{idapro} and \texttt{x64dbg}~\cite{x64dbg} often cause the app to crash when they single-step through function calls to a DLL.
And since each {\app} uses a different set of external DLLs, we could not establish a single workflow to analyze all {\app}s.
Existing tools such as TaintDroid~\cite{enck2014taintdroid} are able to establish the data flow within an application in older Android versions.
However, in native applications designed for Windows and macOS, flow tracing is difficult and sometimes impossible.

% As such, we moved to implement a makeshift information tracing system to follow the microphone data inside the Webex Windows app.
% We started by identifying the source and sink for the information of interest: the microphone generates data that ultimately may be sent to the network.

It is easy to see when an app accesses the hardware (networking and microphone) by monitoring Windows API calls (see Sec.~\ref{sec:methods}), but we are not aware of any tool that can automatically follow data through an entire Windows app. Tracking microphone data after each instruction is not straightforward. Such data initially exists inside of dynamically-allocated memory buffers. Upon each access, this data might move to a new buffer after undergoing a transformation, such as encryption, compression, or encoding. Further, the new buffers may originate from different allocator functions to be stored in the main process's memory image or in an external DLL's memory image.  Race conditions among the threads in Webex compound the difficulty of tracing: all memory accesses at a specific address of the stack require stoppages, logging, and memory analysis, all of which take time to perform.

% Each time an app's microphone processing thread calls a function, that function may allocate a new buffer and copy the microphone data into it.
% The newly-allocated buffer may contain exactly the same data as the raw microphone buffer, or --- more likely --- it may contain some modified version of the original, processed by encryption, compression, encoding or some other algorithm.
% The buffers may originate from \texttt{malloc}, \texttt{new}, a custom memory allocator, or some combination.
% The heap that contains the new buffer may be 
% And the only clue that the buffer exists at all is a pointer on the microphone thread's stack.
% Each instruction executed in the microphone thread potentially creates a new buffer containing audio-derived data, making it effectively impossible to track manually. 

However, we do not necessarily need to show a linkage between every successive subroutine that handles microphone data in a {\app} to demonstrate that audio-derived data leaves on the network.
We can already dynamically trace the audio into the app.
We need to show that data from that buffer leaves our machine and is transmitted to a Webex server.

To design such a system, we first map all of the outgoing traffic from Webex. The most efficient way of doing so is to use the Microsoft Network Monitor (MNM). We observe Webex's network traffic using the MNM while the app is muted and unmuted, and we notice a set of packets that are periodically going to a user metrics Cisco server. Now that we have our packets, we need to ensure that the audio buffer within Webex is accessed in the muted state.

%%%Kassem: I incorporated the below in other places. 
% This can be accomplished when attaching to the app using \texttt{x64dbg}~\cite{x64dbg}, a real-time binary decompiler for Windows apps, and API monitor~\cite{apimonitor}. We utilize an anti-anti-debugging library called Scylla-Hide~\cite{scyllahide}, which hides the fact an app is being debugged from the Webex.

% The API monitor gives us the address in memory, while the app is running, of the raw audio bytes, and \texttt{x64dbg} allows us to trace Webex's copied audio buffer until that buffer reaches the stack.
% We found that while the app was muted, Webex's audio buffer contained raw audio from the microphone.

Binary tracing on the audio buffer's read/write access using \texttt{x64dbg} always ends up in stack space which thwarts our further tracing.
However, while following the bytes and logging API calls (from Sec.~\ref{sec:analysis}), we noticed that Webex calls encoding libraries which access in some time correlation with the audio bytes. 
We then trace API calls to encryption methods to verify what is happening using the API Monitor. We capture all input arguments and output buffers as a log file from these calls while the user was muted during a Webex meeting.
The log contains timestamps, input parameters to the API call, and the resulting output buffer.
With the results of the function logged, we compare the encrypted buffer to network traffic leaving the machine and notice a one-to-one match between the encrypted bytes (from Wireshark) and the data sections of network packets from Webex.
Consequently, we link the data regions of outgoing user metrics packets to our post-encrypted output buffers. Upon observing the input, we notice that the input arguments in these cases are in plain-text where detailed data is compressed using base64 encoding.
Decoding the input arguments revealed the packet content to be a JSON structure\footnote{An example of such a structure is here: \url{https://osf.io/szd4x/}}, which contains audio-derived data and other data elements.

% The audio-derived data in question were three attributes, audioMaxGain, audioMeanGain, and audioMinGain.
% These values were the max, mean, and min of gain calculations, done by Webex, of the audio over the course of minute intervals.
% The final task to verify that these values are audio-derived data, we performed correlation experiments of these values compared to raw power levels of selected input audio.

%Ultimately, data tracking through a {\app} is very labor intensive.

% Further, many {\app}s doubly-encrypt the data they send over the network port: the first level of encryption happens inside the {\app}, and the second level happens in the operating system as the packet is prepared for transmission over SSL/TLS\footnote{Secure Socket Layer/Transport Layer Security}.
% After the outgoing data is encrypted within the {\app}, the buffer that encodes the binary data changes, making it impossible to search the {\app}'s memory image for the raw audio data.

% Since decrypting network traffic generated by {\app}s is not feasible, we instead used API hooks to intercept plaintext on its way to the network port before encryption.
% The modules we instrumented are highlighted with bold outlines in Fig.~\ref{fig:Data_Flow}.
% Intercepting traffic is possible on Cisco Webex because it does not encrypt audio data before sending it to the OS. Webex hands over network traffic to the Windows API in plaintext form, which is subsequently encrypted by the operating system's internal SSL implementation.
% See Appendix~\ref{sec:windowsapi} for details about the API functions we hooked. 

 \begin{figure}[t]
     \centering
         \includegraphics[width=\columnwidth]{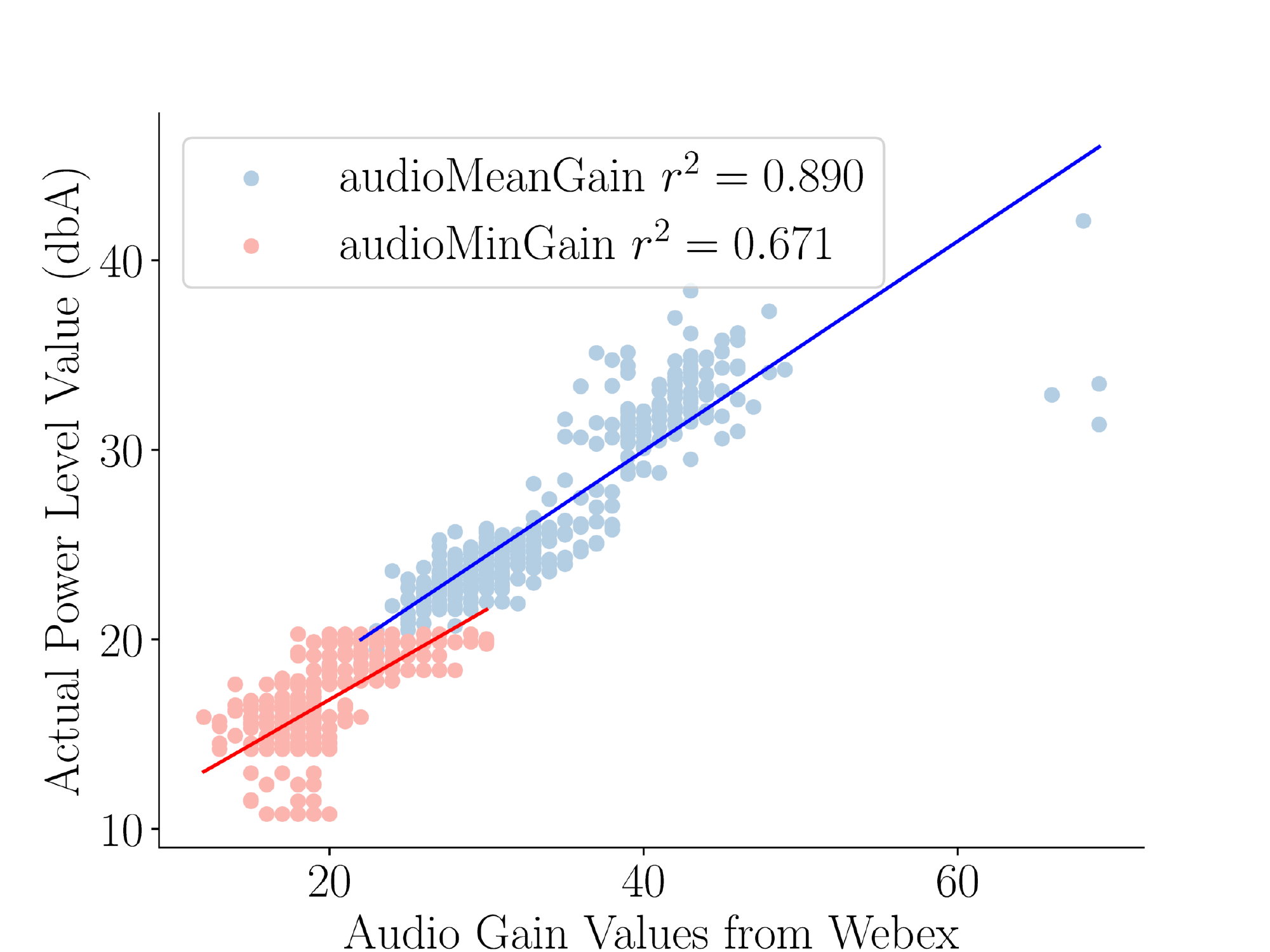}
        %  \caption{Music Mode Enabled
        \caption{Correlation between audio gain reported by Webex and input audio signal power level (in dbA) when noise removal mode is enabled. Although we cannot observe the raw audio while muted, the statistics reported by Webex leak information about the user's background noise.}
        \label{fig:noise_mode_scatter}
\end{figure}

% \vspace{-10mm}

\subsection{Findings for Traffic Interception}
\label{sec:webex_finding}

% \begin{figure}[t]
%      \centering
%          \includegraphics[width=\columnwidth]{PoPETS2022/Figures/correlation_scatter_plot_music_mode.pdf}
%         %  \caption{Music Mode Enabled}
         
%         \caption{Correlation between audio gain reported by Webex and input audio signal power level (in dbA) when music mode is enabled.}
%         \label{fig:music_mode_scatter}
% \end{figure}

The data we capture from the API hook is a JSON array with unencrypted and unobfuscated attribute names such as: \texttt{audioMaxGain}, \texttt{audioMeanGain}, \texttt{audioMinGain}, and many others. These JSON arrays are transmitted by Webex once per minute to \url{https://tsa3.webex.com}, a telemetry server, while the user is muted. The names of these attributes suggest that the JSON array contains audio-derived statistics, most probably connected to the automatic gain control employed by Webex. Our aim is to further analyze the attributes to understand the relationship between the recorded audio levels and these attributes values when the microphone is muted.

Webex has two microphone modes: music mode and noise removal mode (the default mode). As the name suggests, noise removal mode refers to Webex removing background noise in real-time while the user is speaking.
Music mode, on the other hand, transmits audio as the microphone hears it. We perform a small-scale experiment to study whether the audio attributes from Webex network traffic are correlated with the input audio for both microphone modes. We play episodes of the U.S. TV shows ``Friends'' and ``The Office'' into a microphone during a Webex meeting while muted. To isolate environmental factors, we feed the audio from the TV shows directly into the Webex meeting through a virtual microphone interface. We repeated each experiment for both microphone modes.

We partition each audio file (corresponding to an episode) into a set of one-minute windows. We then compute the maximum and average magnitudes for each window to report their correlation with \texttt{audioMinGain} and \texttt{audioMeanGain}. Note that the \texttt{audioMinGain} value would correspond to the maximum observed audio level because it requires less gain control. Further, the minimum and mean values depend the most on the input audio. On the other hand, the maximum depends more on the input device and the amount of silent moments, which are random in each episode.

Fig.~\ref{fig:noise_mode_scatter} depicts the correlation between the estimated power levels and measured gain values for noise removal mode. As evident from the figure, the measured and estimated values exhibit high correlation; the correlation with the mean gain is higher as it is a more robust metric to window shifts. Note that we do not have access to the source code when computing the gain values, so a perfect correlation is unlikely. This correlation is slightly lower than that of the music mode (Fig. \ref{fig:music_mode} in Appendix \ref{sec:music_mode_appendix}), implying that the noise removal changes the input audio. Still, the measured \texttt{audioMinGain} and \texttt{audioMeanGain} are representative of the audio levels.

%% file: PoPETS2022/5_classification.tex
\subsection{Classification of Background Activities}
\label{sec:bg_classify}
% \revreq

We established that Webex accesses the microphone while muted and sends audio statistics to their servers. Further, this data is highly correlated with the energy level received at the microphone, and appears to be indicative of the activity happening in the background. The logical question that follows is: \emph{is there a potential of learning the user background activities from audio statistics sent to Webex's servers?} In the following, we describe how these statistics can fingerprint the user's background activities, when they are muted.

We analyze the inference of information from user's Webex telemetry traffic while being muted. For each one-minute window, this information contains three values that change relatively : mean, min, and max audio gains. An entity with access to this information, such as Webex's cloud service or any adversary able to view this traffic in transit, can perform this analysis to infer what activity is occurring in the user's environment.

\subsubsection{Data Collection}
We focus on the background activities from our user study of Sec.~\ref{sec:user_study}. In Fig.~\ref{fig:codes_q2}, we highlight twelve activities that happen in the user's background. Out of these activities, we do not consider: (1) silent and physical activities as they do not result in gain changes, (2) bathroom as it is unlikely that the user's microphone will pick up bathroom noises, (3) street noise as it does not represent a private activity, and (4) diverse noise such as TV shows which may contain all of the classes in a single 30-minute instance. As such, our objective is to identify whether the gain values can fingerprint six types of activities: (1) music playing, (2) cooking or eating, (3) people talking, (4) animal sounds (especially dog barking), (5) keyboard typing, and (6) cleaning.

To simulate the real-world environment with specific background activities, we choose multi-hour long ASMR YouTube videos that consist of single background activity. Each video is different such that the videos are produced by different people (YouTube users) doing the same task. The purpose of selecting the videos in such a way is to minimize the effects from the recording environment.  We play each video over the air through a Webex meeting, while muting the microphone, and log the extracted gain values. 

Our data collection consists of two Windows 10 machines. The first machine plays the videos using its speaker and hosts the meeting for the other machine. The other machine runs a Webex meeting client (without any other app running). One machine is equipped with a Logitech QuickCam Pro 9000 while the other uses a Logitech C920S Pro HD 1080p webcam for microphone input. Both machines then join the same meeting room and collect data simultaneously; on both, we mute the microphone, turn off the camera, and keep the default microphone settings. 

We place the machines in a $12ft\times7ft\times10ft$ room. We adjust the distance from the speaker to the two microphones and generate multiple datasets based on the varying distances. Webex only allows for meetings to last for 24 hours. For each Webex meeting, we can extract around 1440 data points, stamped with the corresponding label. Each data point corresponds to three features: \texttt{audioMaxGain}, \texttt{audioMeanGain}, and \texttt{audioMinGain}, representing three user metrics values from one minute of audio. In summary, we performed data collection over the course of two months, yielding over 180 hours of data points.

\begin{figure}[t]
     \centering
    \includegraphics[width=0.9\columnwidth]{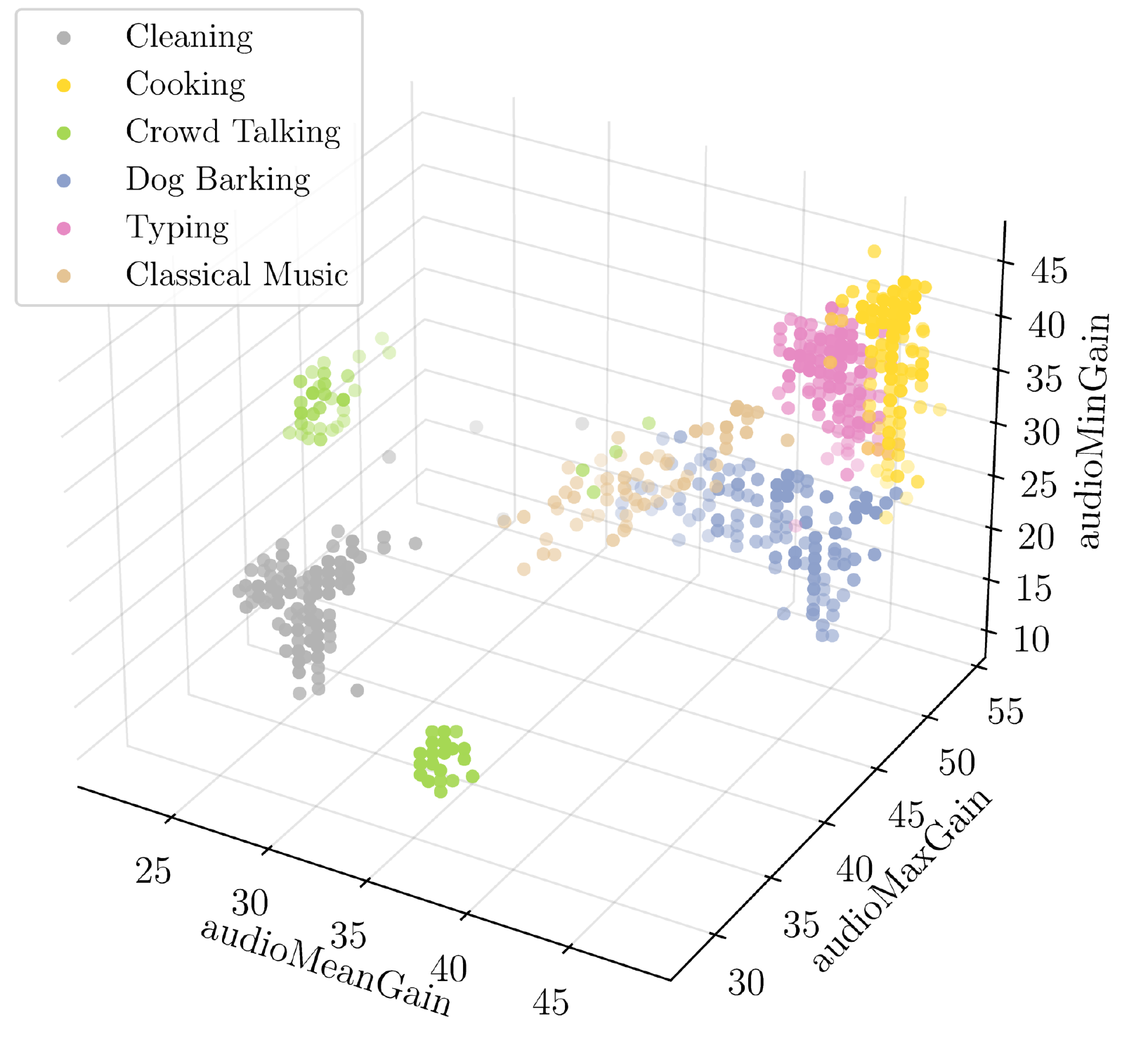}
         \caption{ Clusters of audio statistics data color coded by background activity type. Clusters are visually separable. }
         \label{fig:bg_visual}
\end{figure}

We visualize the distribution of the six background activities in Fig.~\ref{fig:bg_visual}. This figure shows that it is feasible to fingerprint background activities by analyzing the extracted gain values from Webex. Each activity exhibits relatively consistent and distinguishable gain values, despite sampling diverse videos to represent each activity.

\subsubsection{Classifier Training}
We design a classifier to highlight how the background activities can be fingerprinted based on the observed gain values. In what follows, we describe how we curate the data for this classifier, how we design and train the classifier, and the results of the classification.

\paragraph*{Data Preprocessing}

We split the YouTube videos into a development set (training and validation) and an evaluation set. The two sets have no overlaps in the videos. We set the distance from the microphone to speaker as 10~cm, 25~cm, and 50~cm for both sets of videos, whereas we added an extra distance condition, 100~cm, for the evaluation set.

Table~\ref{tab:ds_dist} shows the data distribution for the development and evaluation sets. We split the development set into a training set (80\%) and validation set (20\%) for hyper-parameter tuning.  We split the evaluation set into two subsets to study the effect of distance. The first evaluation subset is collected at distances of 10~cm, 25~cm, and 50~cm between the speaker and microphone. The second subset is collected at a distance of 100~cm; the data in the second evaluation subset has no overlap with the development set in terms of distance and source videos.

For a $t$-minutes long YouTube Video, we extract $t$ data points; each data point is assigned the same label derived from the title of the video. To accommodate videos of varying lengths, we limit the input to the classifier to clips of length $n$. Thus, we apply a sliding window with length $n$ to each window and set the moving stride to be 1. We define each clip as: 
\begin{equation}\label{eq:data_clip}
    \textsc{Clip} = 
    \begin{bmatrix}
        \mathrm{max}_i & \mathrm{max}_{i+1} &  ... & \mathrm{max}_{i+n-1}\\
        \mathrm{mean}_i & \mathrm{mean}_{i+1} &  ... & \mathrm{mean}_{i+n-1}\\
        \mathrm{min}_i & \mathrm{min}_{i+1} &  ... & \mathrm{min}_{i+n-1}\\
    \end{bmatrix},
\end{equation}
where  $\mathrm{max}_i$ represents the \texttt{audioMaxGain} for the $i^{th}$ minute in the window. 

\begin{table}[t]
\centering
\begin{tabular}{lccccc}
Class             & Train & Val & Eval1 & Eval2 \\ \midrule
classical music  & 168      & 43       & 379      & 184      \\
cooking/eating & 500      & 126       & 486      & 169      \\
crowd talking    & 656      & 164      & 1191     & 568      \\
dog barking      & 408      & 103      & 726      & 691      \\
keyboard          & 359      & 90       & 1324     & 580      \\
vaccume/cleaning  & 544      & 136      & 668      & 572      \\
total (minutes)   & 2637     & 660      & 4774     & 2764     \\ 
\end{tabular}
\caption{Dataset distribution, development set (training and validation) and evaluation set (subset 1 and 2). }
\label{tab:ds_dist}
\end{table}

\paragraph*{Model Design and Training}

We train a supervised multi-class classifier to distinguish background activities given clip data of length $n$. Similar to Schuster et al. ~\cite{schuster2017beauty}, we use a Convolutional Neural Network (Fig.~\ref{fig:bg_model} in Appendix); the network consists of two 1-dimensional convolution layers, flatten layer, three dense layers, and a softmax layer (of size 6). The design of the convolutional layer takes into account feature and temporal correlations.

We train the network using an Adam optimizer with a cross-entropy function as the loss calculation function. We set the learning rate to 0.001 and initialize model parameter weights in a random uniform distribution. As the total length of the training set and validation is around 3000, we evaluate different batch sizes: 50, 500, 1000, 1500, and 3000. We utilize early stopping to prevent over-fitting. Because the dataset is imbalanced, we calculate the precision separately for each class and compute the average precision score weighted by their proportion in the validation dataset. Then we use the weighted average of precision score and accuracy of all classes as the early stopping criterion.  We select the best-performing epoch index and batch size to train the optimal classifier for each window length. 

We train the network on windows of size $n$: $3,5,7,10$. Comparing the performance of 4 optimal classifiers for each window length, we observe that $n=7$ (96.13\% precision on validation set) outperforms windows $n=3$ (92.26\%)  and $n=5$ (92.98\%), in terms of the accuracy score and precision score of the validation set. We achieve 96.90\% with windows length $n=10$ but we remove it in case of over-fitting.

\subsubsection{Classification Results}

\begin{figure*}[t]
     \centering
     \captionsetup[subfigure]{justification=centering}
     \begin{subfigure}[b]{0.3\textwidth}
         \centering
         \includegraphics[width=\textwidth]{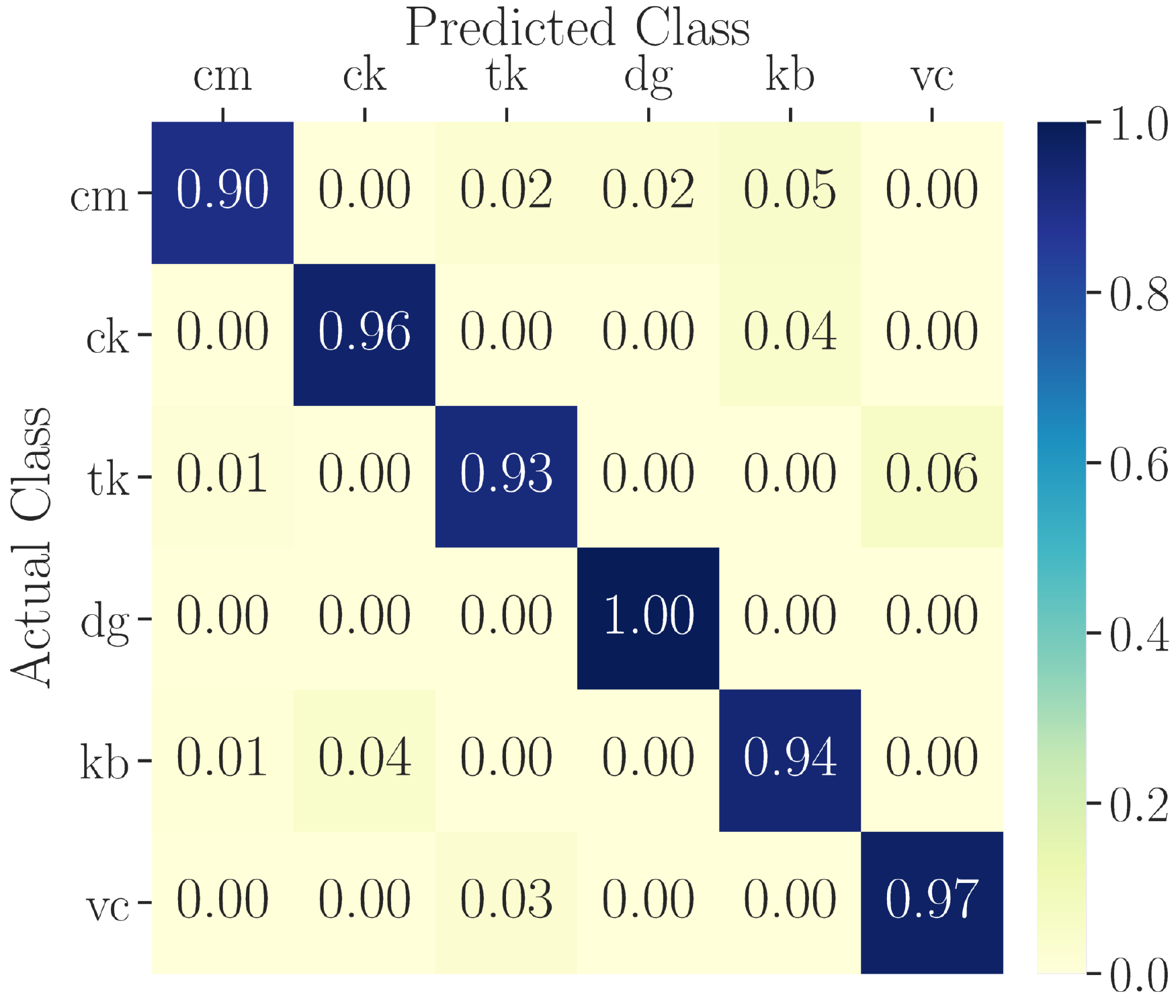}
         \caption{Validation Set}
         \label{fig:valdset}
     \end{subfigure}
     \hfill
     \begin{subfigure}[b]{0.3\textwidth}
         \centering
         \includegraphics[width=\textwidth]{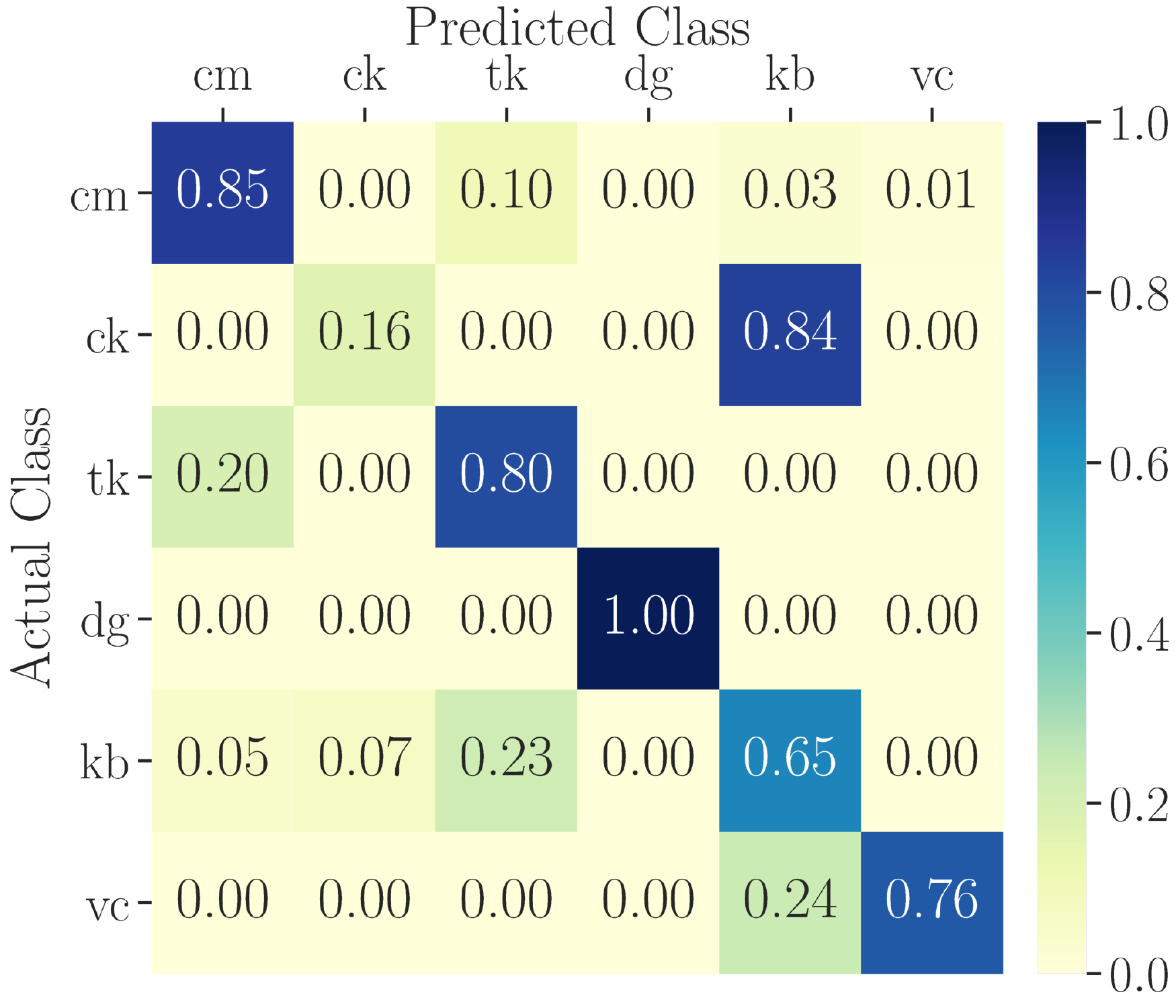}
         \caption{Evalset1}
         \label{fig:eval1}
     \end{subfigure}
     \hfill
     \begin{subfigure}[b]{0.3\textwidth}
         \centering
         \includegraphics[width=\textwidth]{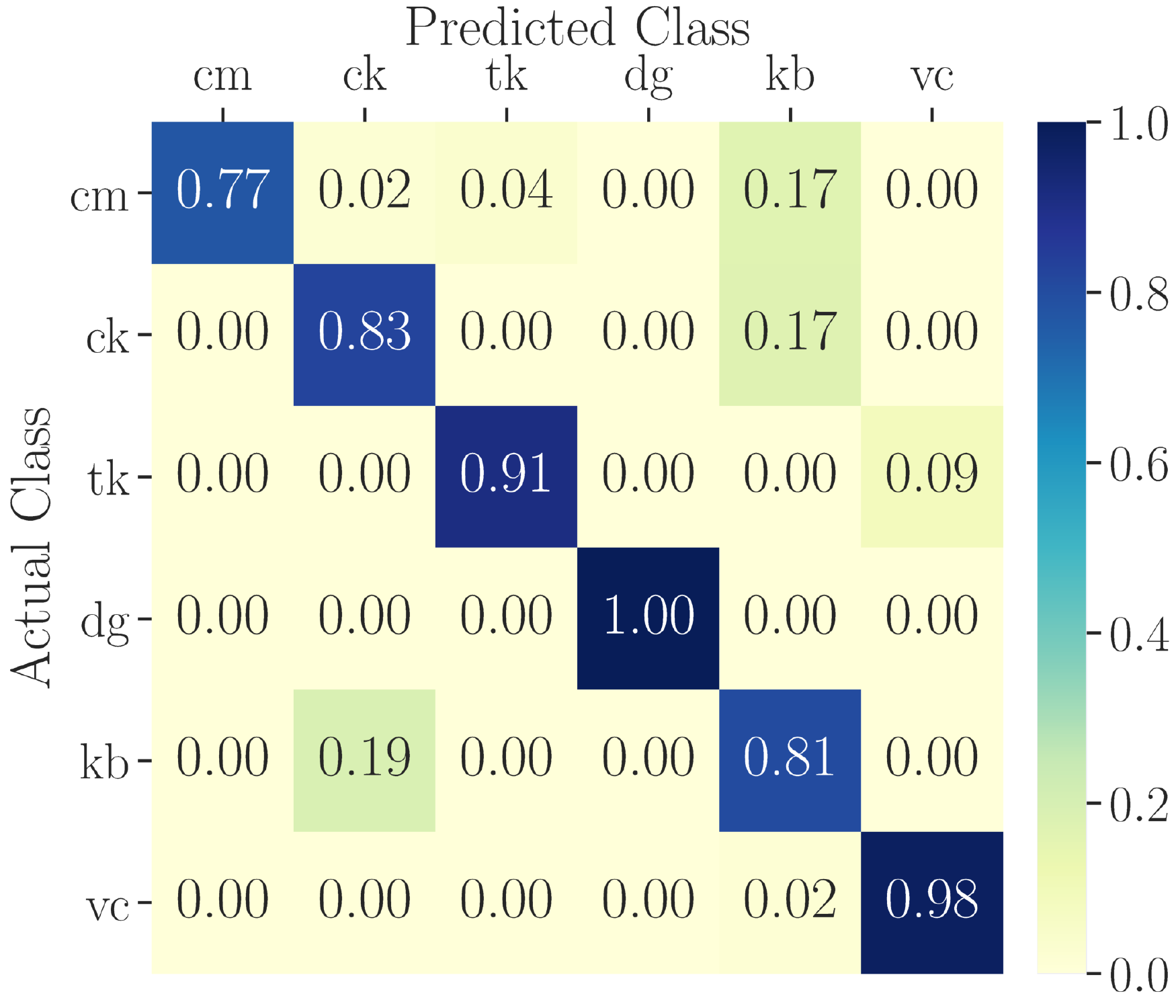}
         \caption{Evalset2}
         \label{fig:eval2}
     \end{subfigure}
        \caption{Background activity classifier performance with window length = 7. The six classes include Classical Music (cm), Cooking or eating (ck), Talking (tk), Dog Barking (dg), Keyboard (kb),  Vacuum or Cleaning (vc).}
        \label{fig:cofm_bg}
\end{figure*}

We present the per-class performance of classifier with window lengths 3 and 7 in Fig.~\ref{fig:cofm_bg_win3} and Fig.~\ref{fig:cofm_bg}. 
For window size of $n=7$, we achieve 77.75\% macro accuracy on evaluation set 1 and 89.03\% macro accuracy on evaluation set 2. 
The average of per class precision for evaluation set 1 is 73.07\% while that is 87.47\% for evaluation set 2. Note that evaluation set 2 is collected with 100~cm microphone to speaker distance; our results suggest that the volume level and video content do not considerably hurt the classifier performance.

For window size of $n=3$, we achieve 78.70\% macro accuracy on evaluation set 1 and 78.48\% macro accuracy on evaluation set 2. 
The average of per class precision for evaluation set 1 is 79.35\% while that is 84.35\% for evaluation set 2.
Both classifiers follow our early stopping criteria and achieve high performance on evaluation sets. This performance indicates that, even with three-minutes worth of measurements, it is possible to infer the ongoing background activities.

For both window sizes, dog barking, crowd talking, and cleaning show high precision as well as accuracy on both evaluation datasets. 
Some music and people talking samples are misclassified as keyboard typing on evaluation set 1 and 2 respectively, while cooking and eating shows a lowest performance among the six background activities.
On both evaluation sets 1 and 2, ``cooking or eating'' data points are severely mingled with ``keyboard typing'' as both classifiers cannot accurately classify these two classes at the same time. We discuss this aspect in Sec.~\ref{sec:cf_disc}.

Finally, we test whether Webex's noise-canceling feature affects the statistics reported in log packets. The results are nearly identical with noise-canceling disabled or enabled.
However, there is a difference between the logged gain values from Webex when alternating between the music and noise-removal modes.
Therefore, we only collect and present results based on data collected with noise-canceling enabled --- the default setting --- through our entire classification process.

%%% tone down

Our classifier performs well on both evaluation sets in under various kinds of background noise, recording environments and volume levels.
The gain values logged by Webex and sent to its cloud server can be used to distinguish multiple types of background activity. 

%Comparing the classifiers' performance on both evaluation sets, we conclude that our classifier yield a good performance against multiple video contents, recording  environments, and volume levels.

%% file: PoPETS2022/6_discussion.tex
\section{Discussion}
\label{sec:discussion}
In the following, we discuss some of the limitations with our methodology. We also discuss possible mitigation strategies, including an improved OS-level permission model and user education.

\subsection{Limitations of the Study}
\label{sec:cf_disc}

Using live binary analysis tools, we developed a technique to trace incoming audio data from the microphone driver to the operating system's socket API; our methods are in compliance with each app's Terms of Service (ToS).
We conducted a thorough evaluation of the Webex native Windows app, demonstrating that we could distinguish a variety of background activities that were most commonly reported in our user study.
We discuss limitations in (1) our binary analysis techniques, (2) our dataset and (3) our background activity classifier.

The first limitation is that our binary analysis technique does not easily generalize to other apps because different {\app}s have different mechanisms for preparing and encrypting network traffic.
Many of the apps we studied encrypt the outgoing data stream before passing it to the operating system's socket interface, making it impossible to search the binary's memory image for the raw microphone data.
Only in Webex were we able to intercept plaintext immediately before it is passed to the Windows network socket API.

The second limitation is that the findings from the user study might not generalize to the general population. The user study participants are young and educated professionals, who are potentially more tech-savvy than the general population. However, the responses to our questions did not reveal a high level of technical sophistication when describing the operation of the mute button. Fig.~\ref{fig:codes_q3} shows that handful of participants were able to correctly describe the operation of the mute button.

The third limitation is that we collected data for our Webex case study in only one room.
We do not consider the impacts of the speaker's volume level or the room's acoustic properties that may affect the microphone input.
It may be possible to infer a relationship between the room's acoustic properties and the audio statistics that Webex reports using raw audio data acquired while the app is unmuted.

Finally, our classifier targets single background activity at a time, and it does not perform well on all background activities.
Differentiating between multiple sources is potentially possible, however, due to a limited data collection scheme we did not evaluate multiple simultaneous events.
Furthermore, the  ``cooking'' background activity shows a low accuracy score and overlaps with ``keyboard'' data points in Fig.~\ref{fig:bg_visual}.
Poor performance of the cooking class appears to be caused by inconsistent noises that are generated by different cooking activities like grilling, frying, baking, etc.
Another reason for the poor performance is that cooking and typing sound similar at different distances.
Also, our data does not account for noises that are short in duration.
Sounds need to last at least a single minute to create a data point; our techniques cannot evaluate unique but short noises.

\subsection{An OS-Level Mitigation}

To ensure a trustworthy permission model for microphone access in {\app}s, we suggest that operating systems adopt a ``software mute'' feature similar to the one implemented in Chromium and WebRTC.
Under that model, the {\app} calls an API function or syscall in the OS to disable audio traffic flowing from the microphone driver to the app, putting the OS in charge of the microphone data while the app is muted.

The OS's microphone status indicator would serve as an easy and nontechnical mechanism for users to audit {\app}s, ensuring that they use the software mute correctly.
The microphone status indicator should be on only when the {\app} is unmuted and off otherwise.
In our analysis of mute button, we found that the operating system cannot detect the state of an app-controlled mute button, and consequently the microphone status indicator does not correctly reflect whether the {\app} is actively reading data from the microphone driver.
Since the mute functionality is currently implemented in the {\app} instead of the OS, there is no clear policy about how microphone data should be handled during mute that applies to every {\app}.
As we discovered, some apps read from the microphone at a lower data rate during mute, but Webex reads from the mic the same way regardless of mute button status.

An OS-mediated software mute establishes clear rules about when the {\app} should be reading from the microphone, making it clear to the OS when the microphone status indicator should be illuminated and making it clear to the user when the {\app} is reading from the microphone.

\subsection{VCA Privacy Policies}
Few participants in our user study were aware of the data collection or sharing policies of popular {\app}s.
Around 70\% of our participants believe that the mute button blocks the transmission of microphone data or disables the microphone altogether.
{\app} service providers should provide detailed definitions of data collection scenarios rather than generic statements about how they collect data about their users. All VCAs actively query the microphone when the user is muted, and they might have legitimate purposes. For example, Zoom alerts the user when they try to speak with their microphone muted. The privacy policies of these services need to be explicit about microphone access, which is not currently the case.

We analyzed the privacy policies of the {\app}s from Table 2 to understand how do they describe their privacy practices. Other than Google~\cite{google_privacy_policy}, no privacy policy makes an explicit mention to the mute button and how microphone data is accessed when the user is muted. The mention of the mute button in Google Meet's privacy policy refers to the meeting organizer's ability to mute others. Also, the privacy policies are vague about the data collected when the user is running a {\app}.
Some privacy policies, such as Whereby's and Google Meet's, explicitly mention that they do not collect audio data. Other {\app} privacy policies do not mention collecting audio data at all. 
Most policies describe their data collection, in general terms, as ``depend[ing] on the context of your interactions''~\cite{microsoft_privacy_policy}. 
The common reasons that {\app} service providers cite for collecting data are to improve ``app performance''~\cite{microsoft_privacy_policy,cicsoprivacy}, to facilitate ``research''~\cite{microsoft_privacy_policy,whereby_privacy_policy}, and for ``user analytics''~\cite{microsoft_privacy_policy,whereby_privacy_policy,cicsoprivacy}.

Interestingly, Cisco's privacy policy~\cite{cicsoprivacy} mentions audio data in the context of ``types of personal information that [Cisco] may process.''  Cisco's privacy policy is not specific about when the collection is happening and about the purposes of this collection. However, a different privacy datasheet~\cite{webex_prvdatasheet} from Cisco mentions:

\begin{quote}
Cisco Webex Meetings \textbf{\textit{\underline{does not}}}:

Monitor or interfere with your meeting traffic or content.
\end{quote}

Our findings suggest that, contrary to the statement in the privacy policy, Webex monitors, collects, processes, and shares with its servers audio-derived data, while the user is muted. To inform Cisco of our investigation results, we opened a responsible disclosure with Cisco about our findings. As of February 2022, their Webex engineering team and Privacy team are actively working on solving this issue.

%% file: PoPETS2022/8_conclusion.tex
\section{Conclusion}

In this paper, we present the first large scale study of {\app} mute-button privacy.
Our user study shows that users are unaware of Webex listening to their microphone while muted.
We examined all widely used {\app}s and desktop operating systems and pinpointed a potential privacy leakage within Webex.
We discovered that while muted, Webex continuously reads audio data from the microphone and transmits statistics of that data once per minute to its telemetry servers.
Using runtime binary analysis tools, we intercepted unencrypted copies of the telemetry data before it was transmitted.
We used over 180 hours of simulated background noise to build a data set for classification.
Our classifier achieves an 81.9\% macro accuracy on identifying six common background activities using intercepted outgoing telemetry packets when a user is muted.
Operating system vendors can establish a stronger permission model for the microphone by implementing an OS-level software mute.

Our analysis of the {\app}s provide new insight to a user's understanding of the mute button.
We show that Webex transmits audio-derived data while the user is muted.
Counter-measures should be supported by policies and regulations to ensure that users’ private background activities are not monitored.

\section*{Acknowledgements}
This work was supported in part by the National Science Foundation under the awards CNS-1838733, CNS-1942014, CNS-2003129, and OAC-2107020. The authors would like to thank Ahmed Fawaz for observing that some apps on his iPhone keep accessing the microphone, even while muted. His observation inspired the research in this paper.

%% file: PoPETS2022/appendix.tex
\begin{appendix}

\section{Model Architecture}
\begin{figure}[h]
    \centering
    \includegraphics[width = \linewidth]{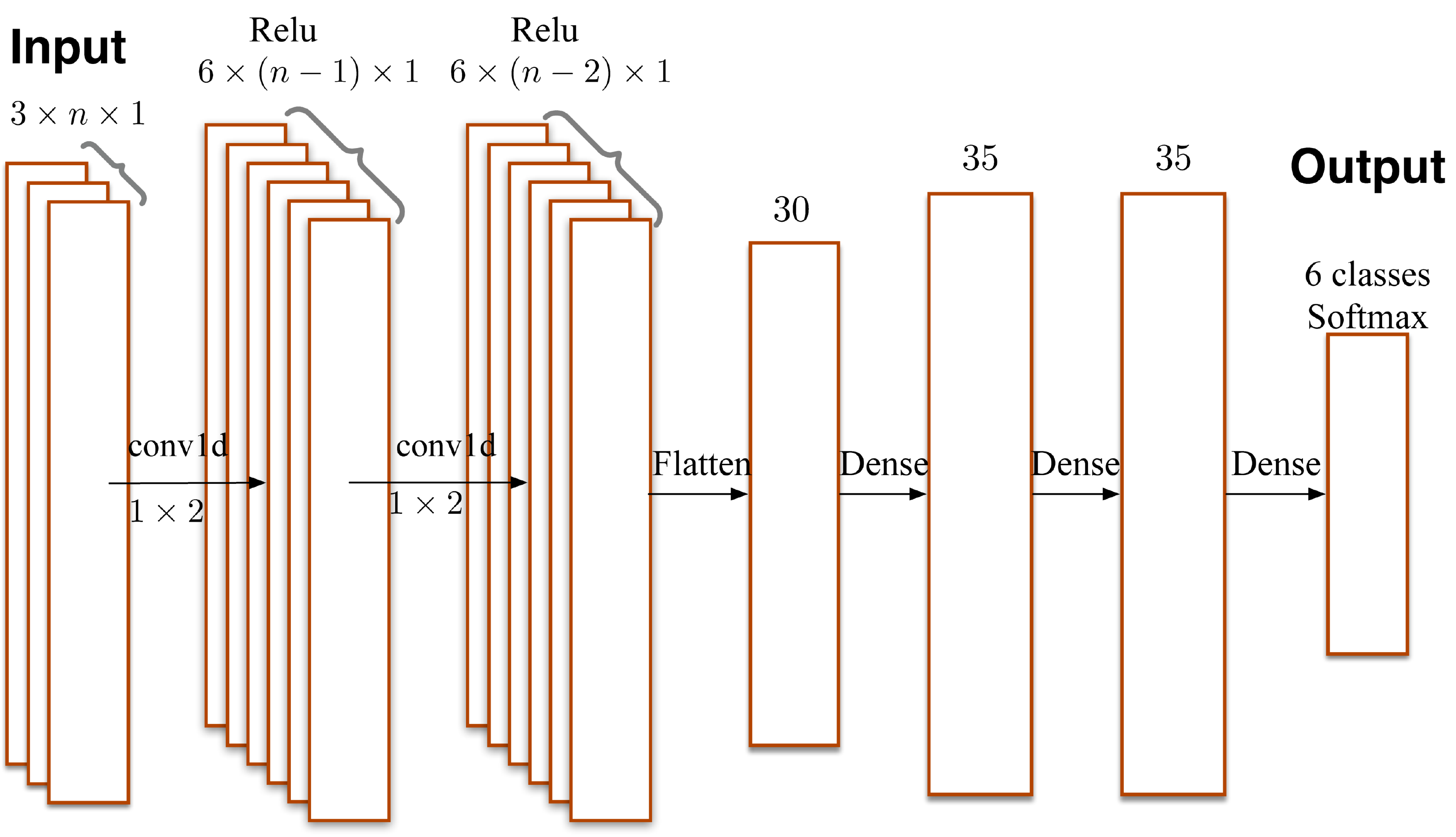}
    \caption{The classification architecture for the background activities.}
    \label{fig:bg_model}
\end{figure}
During training, we observe that batch size affects training speed and performance.
Our classifier is trained with batch size of 500, epoch of 400, window length of 7 and learning rate of 0.001 .

\section{Chromium API}
\label{sec:apiappendix}

Chromium acted as a layer between the operating system and the browser based VCAs.
To verify microphone access we injected our own logging scripts in the source code of Chromium.
Knowing when an app accesses the microphone requires several functions to be monitored, the main functions we observed were:
% -- retrieves the raw microphone data from the operating system.
% Implemented in C++ as a part of the Chromium browser, this function physically retrieves the microphone data from the operating system.
% -- transforms the raw audio into an encoded stream.
%     This function is part of the WebRTC API within the WebKit GTK~\cite{webkitgtk}, the standard hardware interface used by Mozilla and Chromium.
\begin{enumerate}
    \item \texttt{PulseAudioInputStream::ReadData()}
    \texttt{ReadData} indiscriminately reads audio frames from the operating system into a local buffer, regardless of the VCA's mute status (muted or unmuted).
    \item \texttt{opus\_encode\_native()} 
    After receiving a full microphone audio frame from the operating system, \texttt{PulseAudioInputStream::ReadData()} passes that frame to \texttt{opus\_encode\_native()}, regardless of the VCA's mute status.
    \item \texttt{AudioSendStream::AudioSendStream()} -- transfers the encoded audio stream to the web-based VCA.
    It is also a WebRTC API call that executing code can call.
    \texttt{AudioSendStream} only hands the encoded audio data to the VCA if WebRTC's software mute function is disabled.
\end{enumerate}

These three functions outline a general flow of audio data within Chromium in Linux(as of writing this).
Logging important variable's states within these three functions painted an accurate picture of microphone usage while the user was muted.

\section{YouTube Video List}
\label{sec:youtubelist}

Development Set (Training set and validation set) is based on YouTube Video List I. Evaluation set is based on YouTube Video List II.
\subsection{YouTube Video List I}
\begin{itemize}
    \item Dogs Barking for 12 hours - High Quality Sounds: \\ {\small \url{https://www.youtube.com/watch?v=3Go2_VXy1Tg}}
    \item ASMR One Hour of Soothing Grill Sounds – Sizzling Meat: \\{\small \url{https://www.youtube.com/watch?v=NKoJDyKo1Q}} 
    \item Vacuum Cleaner Sound - Extended 10 Hours | White Noise Sounds - Sleep, Study or Soothe a Baby : \\{\small\url{https://www.youtube.com/watch?v=Ms8oZeywjyM&t=7s}}
    \item People Talking Sound effect (10 Hours) : \\{\small\url{https://www.youtube.com/watch?v=y32-rwUr0Nk}}
    \item Baroque Music Collection - Vivaldi, Bach, Corelli, Telemann... : \\{\small\url{https://www.youtube.com/watch?v=ApSoNBu2wt8}}
    \item 10 Hours Typing | Cherry MX Blue Mechanical Keyboard | Gaming Keyboard ASMR : \\{\small\url{https://www.youtube.com/watch?v=h8nmVF0IDCs}}
    
\end{itemize}

\subsection{YouTube Video List II}
\begin{itemize}
    \item ASMR Cooking No talking 5 hours deep relaxation sleeping AD free No ads: \\{\small\url{https://www.youtube.com/watch?v=DoRSCsrKbq8}}
    \item 1 HOUR Barbecue Sound | Soothing Grill Sounds | Sounds : \\{\small\url{https://www.youtube.com/watch?v=va6AOQy8sWM}}
    \item Vacuum Cleaner Sound and Video 3 Hours - Relax, Focus, Sleep, ASMR : \\{\small\url{https://www.youtube.com/watch?v=KilQtE5Nl90}}
    \item Vacuum Cleaner Sound \& Video 2020 Christmas Edition 3 Hours : \\{\small\url{https://www.youtube.com/watch?v=BFNUEVR_Ps8}}
    \item BESTE Baby Einschlafmusik Staubsauger Vacuum Cleaner Sound // 3 Hours // P : \\{\small\url{https://www.youtube.com/watch?v=csHiTtxDmx0}}
    \item 1 Hour of Dog Barking : \\{\small\url{https://www.youtube.com/watch?v=7ej1ur8amCo}}
    \item DOG BARKING 12 Hours Sound Effect : \\{\small\url{https://www.youtube.com/watch?v=fecqn9fnG0s}}
    \item ASMR Typing | Ducky One 2 Mini | Cherry MX Blue (1 HOUR) : \\{\small\url{https://www.youtube.com/watch?v=vlgch5z4y7Y}}
    \item 10 Hours of People talking : \\{\small\url{https://www.youtube.com/watch?v=PHBJNN-M_Mo}}
    \item Anne Pro 1 Hour Keyboard Typing Sounds ASMR (No talking, No music, No mid-roll ads) : \\{\small\url{https://www.youtube.com/watch?v=qMtIOlS_WAo}}
    
\end{itemize}

\section{Music Mode Correlation Results}
\label{sec:music_mode_appendix}

\begin{figure}[!h]
    \centering
    \includegraphics[width = \linewidth]{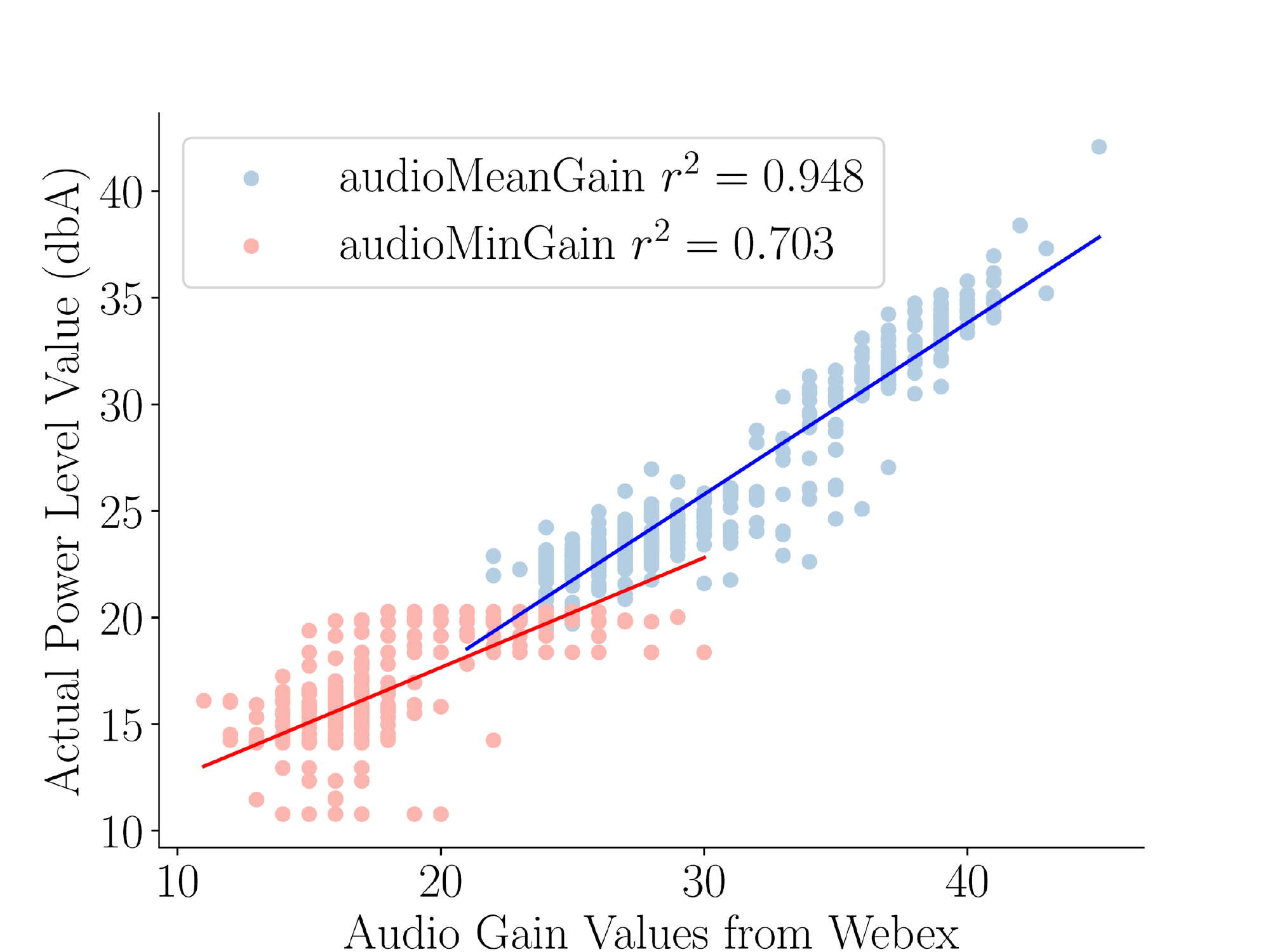}
    \caption{Correlation between audio gain reported by Webex and input audio signal power level (in dbA) when music mode is enabled. Although we cannot observe the raw audio while muted, the statistics reported by Webex leak information about the a user's background noise.}
    \label{fig:music_mode}
\end{figure}

\begin{table*}[!h]
\centering
\captionsetup{justification=centering}
\begin{tabular}{ll}

CodeBook for Q1 & Explanation                                                                      \\ \midrule
No Talk          & No need to talk, so muted; or out of concern, in online classes no need to talk, \\
No interruption  & Do not bother others, do not interrupt others from noise                         \\
Hide Activities  & Hide private activities in the background; hide conversations in background      \\
Generic          & no reason in particular                                                          \\
Comfort          & The participant just feels more comfortable                                      \\ 
\end{tabular}
\caption{Codebook for responses to Survey Question Q1}
\label{tab:code_block_Q1}
\end{table*}

\begin{table}
\centering
\captionsetup{justification=centering}
\begin{tabular}{ll}
\multicolumn{2}{l}{Consolidated codebook - Q2 Activities} \\ \midrule
Music                      & Talking           \\
Dog Barking                & Street Noise                  \\
Watching TV                & Physical Activity             \\
Keyboard                   & Bathroom                      \\
Cooking/eating             & Silent activities             \\
Cleaning/Vacuum            & Online Videos/game            \\
Talking                    & Cleaning/Vacuum               \\ 
\end{tabular}
\caption{Codebook for responses to Survey Question Q2}
\label{tab:code_block_Q2}
\end{table}

\section{Codebooks}\label{sec:codebook}
We present our consolidated codebooks to three open-ended questions (\textit{Q1}, \textit{Q2}, and \textit{Q3}) that are independently generated by two authors in Tables~\ref{tab:code_block_Q1}, ~\ref{tab:code_block_Q2}, and~\ref{tab:code_block_Q3}.

\section{Windows API}
\label{sec:windowsapi}

We can trace the data using the following three methods, which are part of the Windows API DLLs:
\begin{enumerate}
    \item \texttt{BCryptEncrypt} in the \texttt{ncryptsslp.dll} library for encrypting network traffic before sending.
    \item \texttt{IAudioRenderClient::GetBuffer} method in the Windows 10 32-bit Audio interface which fills a local buffer with raw audio data.
    \item \texttt{IAudioRenderClient::ReleaseBuffer} method in the Windows 10 32-bit Audio interface which releases the buffer space acquired in the getbuffer method call.
\end{enumerate}

\begin{table*}[!ht]
\centering
\captionsetup{justification=centering}
\begin{tabular}{ll}

CodeBook for Q3 & Description                                                                                  \\ \midrule
Generic                & A generic description of the mute button                                                                                  \\
Indicator              & visual cue/icon notifying the user of the muting event                                       \\
Block sending          & user experience: block tranmission of audio data to the other clients                        \\
Correct                & the respondent understands the correct operation of mute button                                             \\
Disable Access               & The respondent mentions microphone is disabled or cut when mute button is clicked                   \\
Suspicious             & The respondent suspects the app keeps recording their voice after they apply the mute button \\
Sound detection        & The respondent mentioned the app notify them of possible speaking when muted.                \\ 
\end{tabular}
\caption{Codebook for responses to Survey Question Q3}
\label{tab:code_block_Q3}
\end{table*}

The BCryptEncrypt function is the method that some VCAs executes right before they send a packet over the network.
After this method is executed, Wireshark captures the post-encrypted packet generated from the BCryptEncrypt function as it leaves the machine.
Thus, being able to capture calls at the method before sending the packets grants us unencrypted network traffic.
The GetBuffer method fills a local array in the app's memory space with raw audio data. 
Using the argument's address, we can follow each call and verify if the audio buffer that an app has is changing even while the user is muted.
The ReleaseBuffer method tells us how many frames the app filled their own local buffer with, which gives us a good length of what the app is seeing.
Examining the data we extracted from these methods we can build a dataset that, with confidence, observes audio data from the microphone to the network.

\section{Window Length 5 and 10}
\begin{figure*}[!ht]
     \centering
     \captionsetup[subfigure]{justification=centering}
     \begin{subfigure}[b]{0.3\textwidth}
         \centering
         \includegraphics[width=\textwidth]{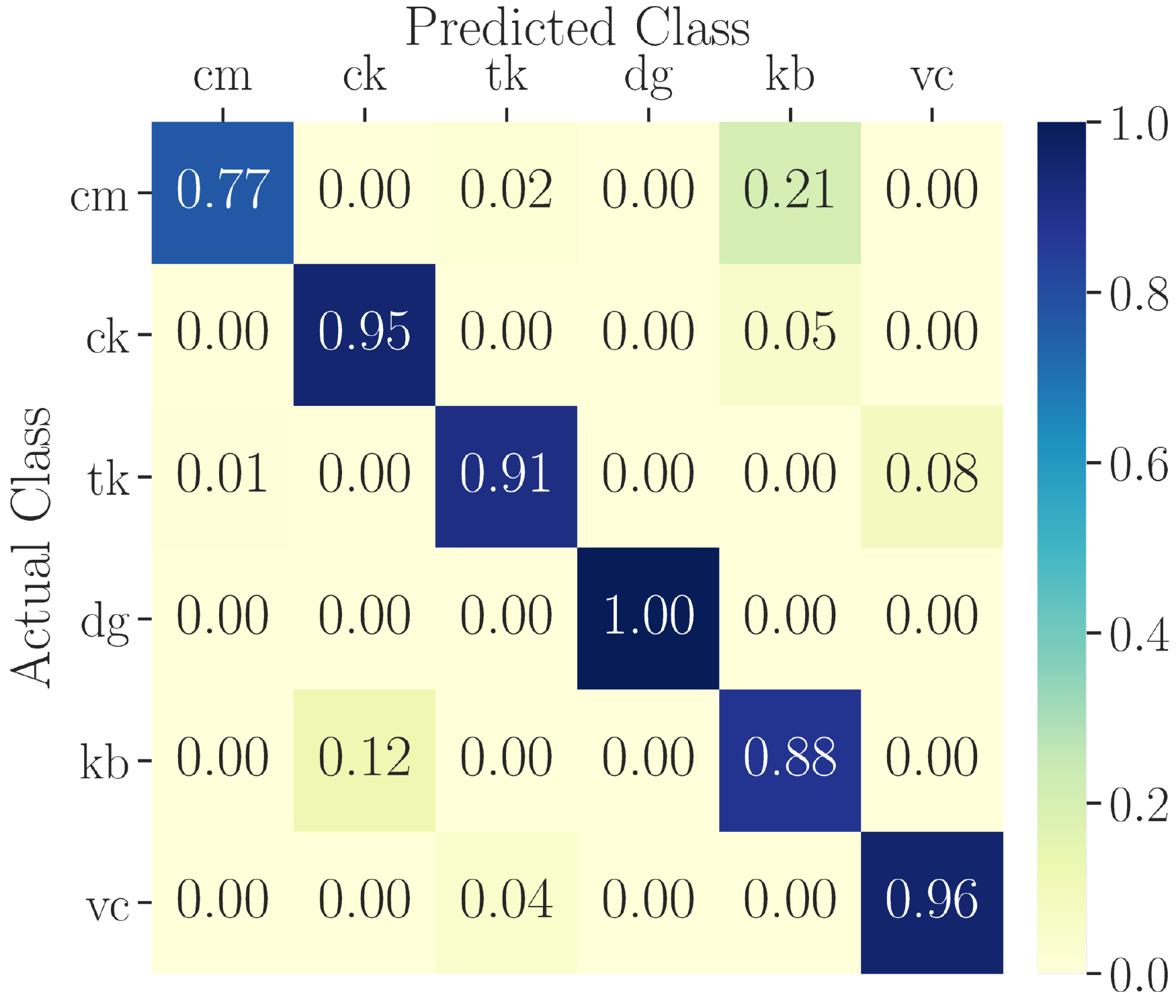}
         \caption{Validation Set}
         \label{fig:valdsetwin5}
     \end{subfigure}
     \hfill
     \begin{subfigure}[b]{0.3\textwidth}
         \centering
         \includegraphics[width=\textwidth]{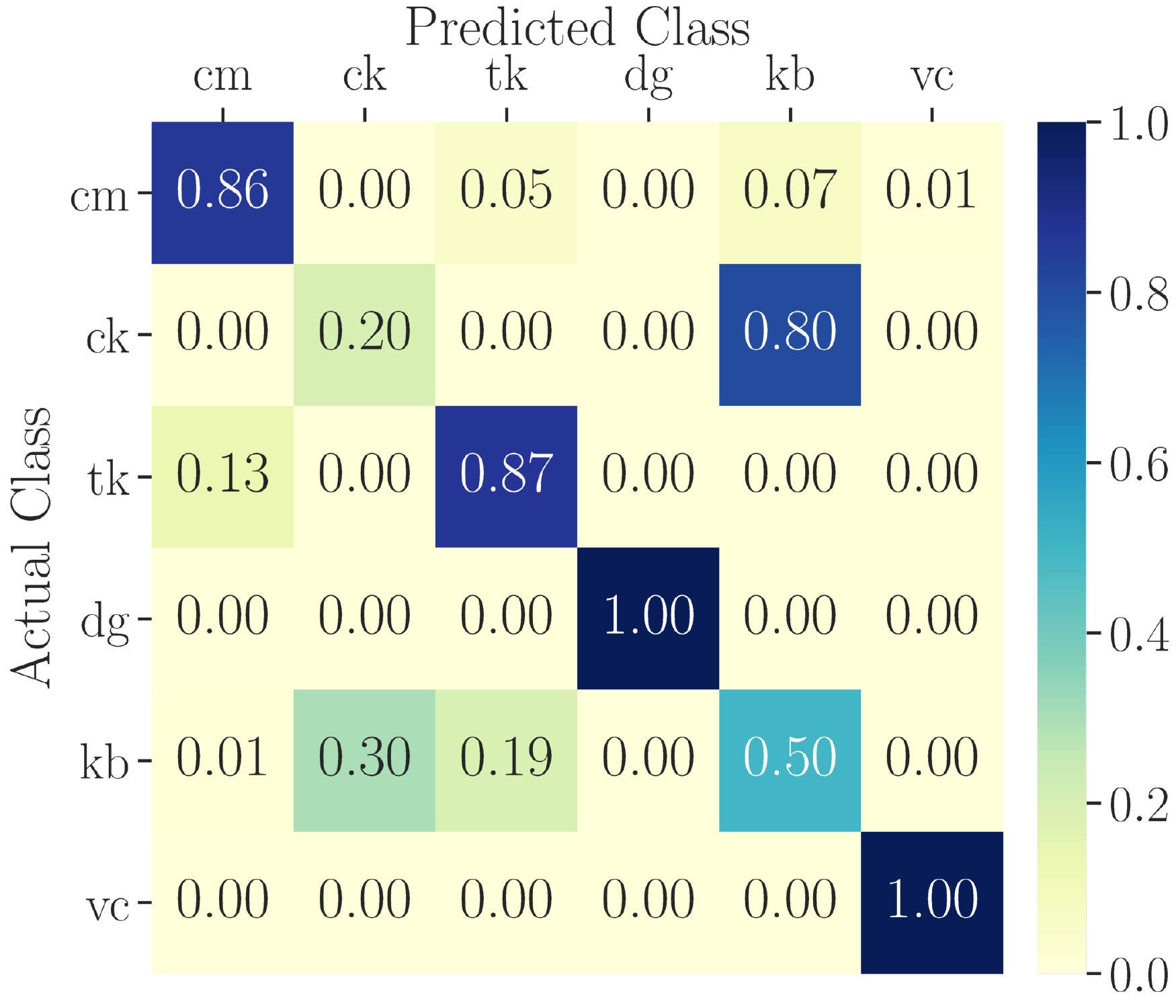}
         \caption{Evalset1}
         \label{fig:eval1win5}
     \end{subfigure}
     \hfill
     \begin{subfigure}[b]{0.3\textwidth}
         \centering
         \includegraphics[width=\textwidth]{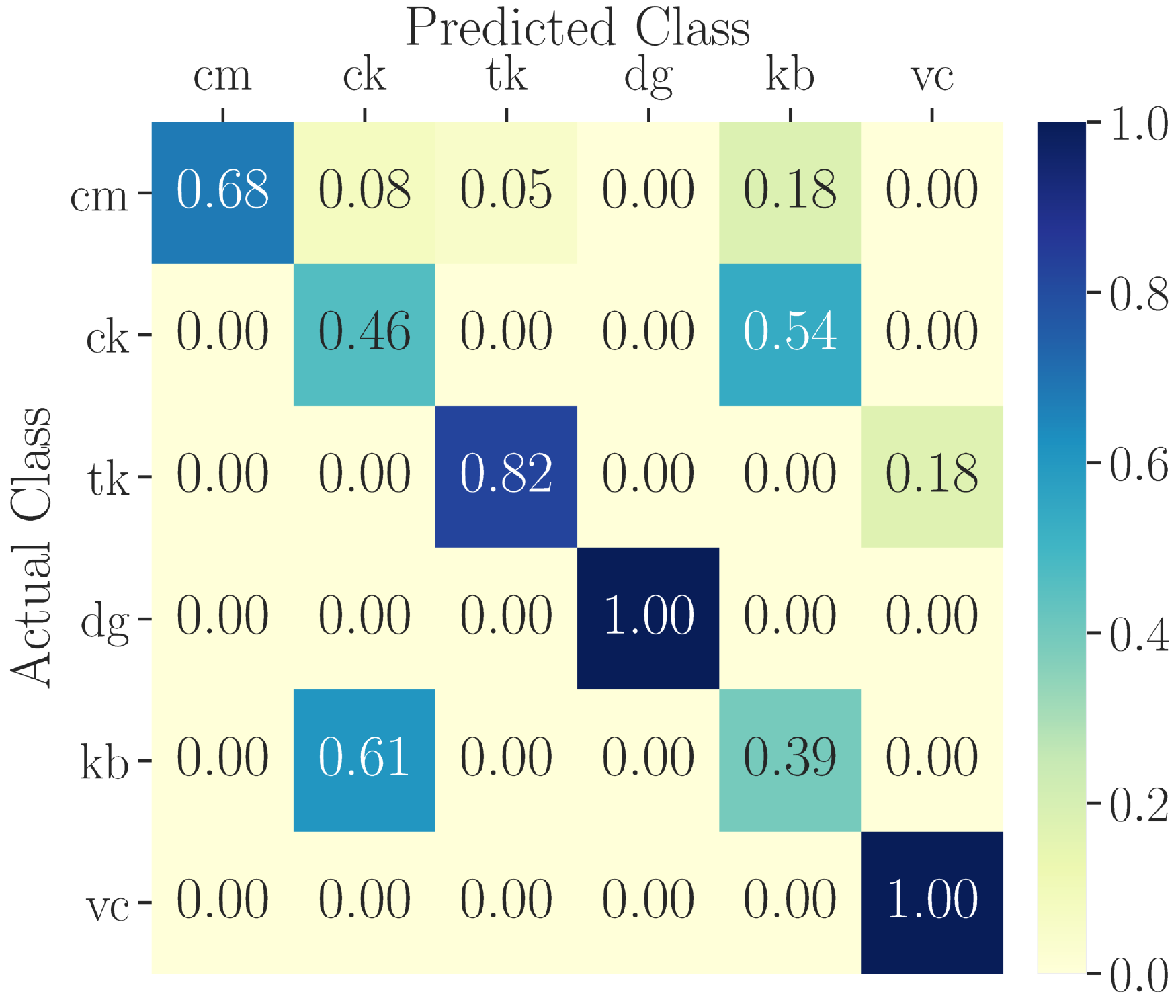}
         \caption{Evalset2win5}
         \label{fig:eval2_5}
     \end{subfigure}
        \caption{Background activity classifier performance with window length = 5. The six classes include Classical Music (cm), Cooking or eating (ck), Talking (tk), Dog Barking (dg), Keyboard (kb),  Vacuum or Cleaning (vc).}
        \label{fig:cofm_bg_win5}
\end{figure*}

\begin{figure*}[!ht]
     \centering
     \captionsetup[subfigure]{justification=centering}
     \begin{subfigure}[b]{0.3\textwidth}
         \centering
         \includegraphics[width=\textwidth]{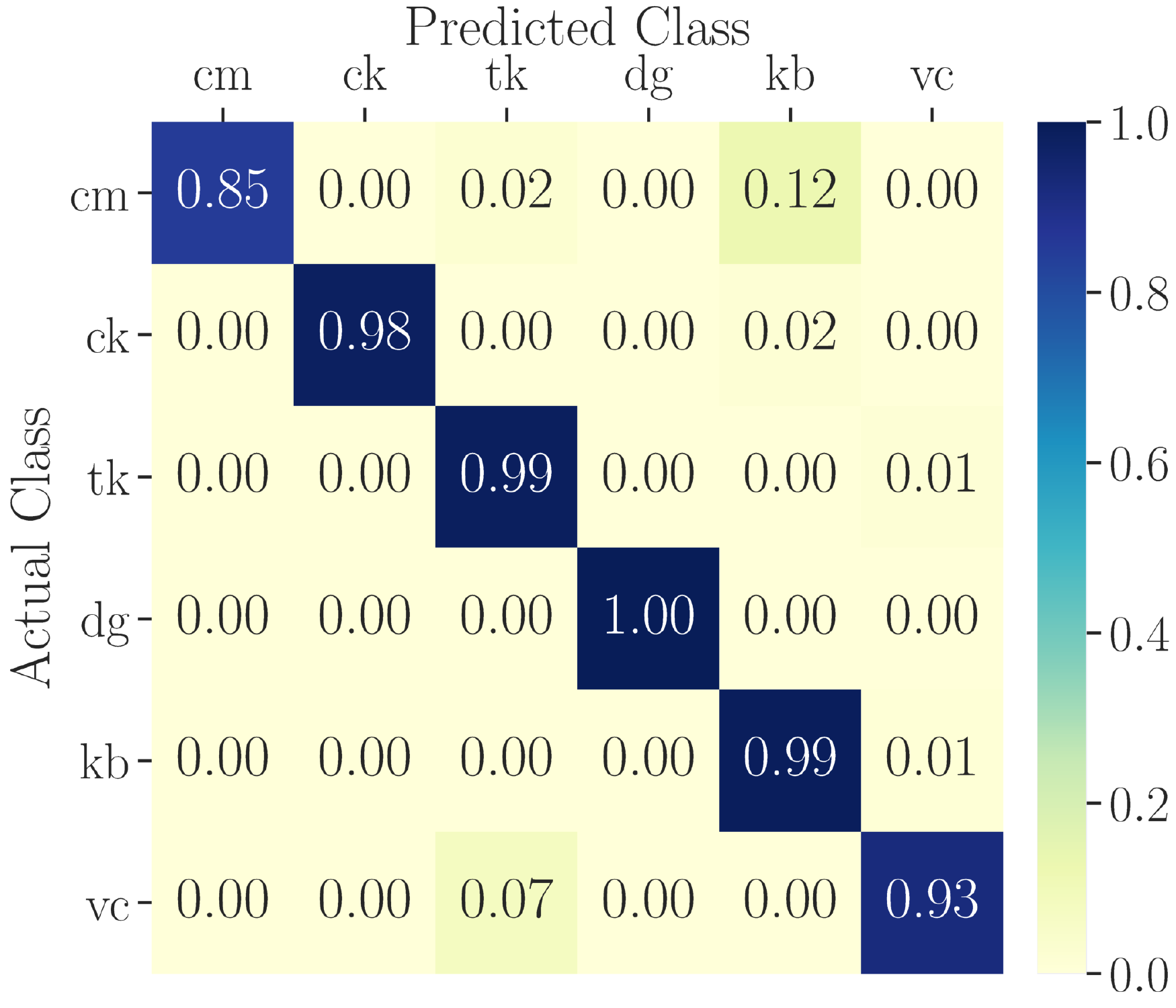}
         \caption{Validation Set}
         \label{fig:valdsetwin10}
     \end{subfigure}
     \hfill
     \begin{subfigure}[b]{0.3\textwidth}
         \centering
         \includegraphics[width=\textwidth]{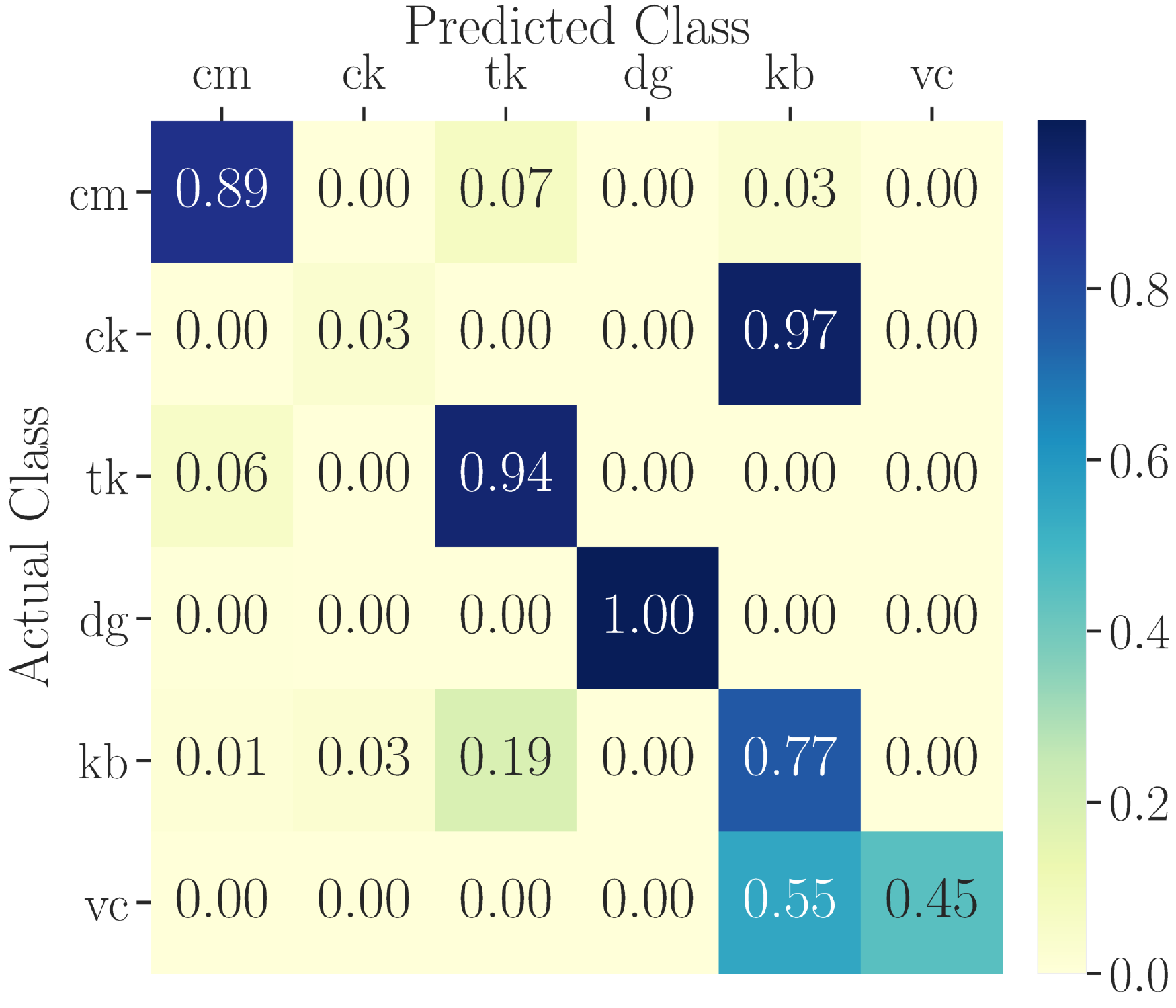}
         \caption{Evalset1}
         \label{fig:eval1win10}
     \end{subfigure}
     \hfill
     \begin{subfigure}[b]{0.3\textwidth}
         \centering
         \includegraphics[width=\textwidth]{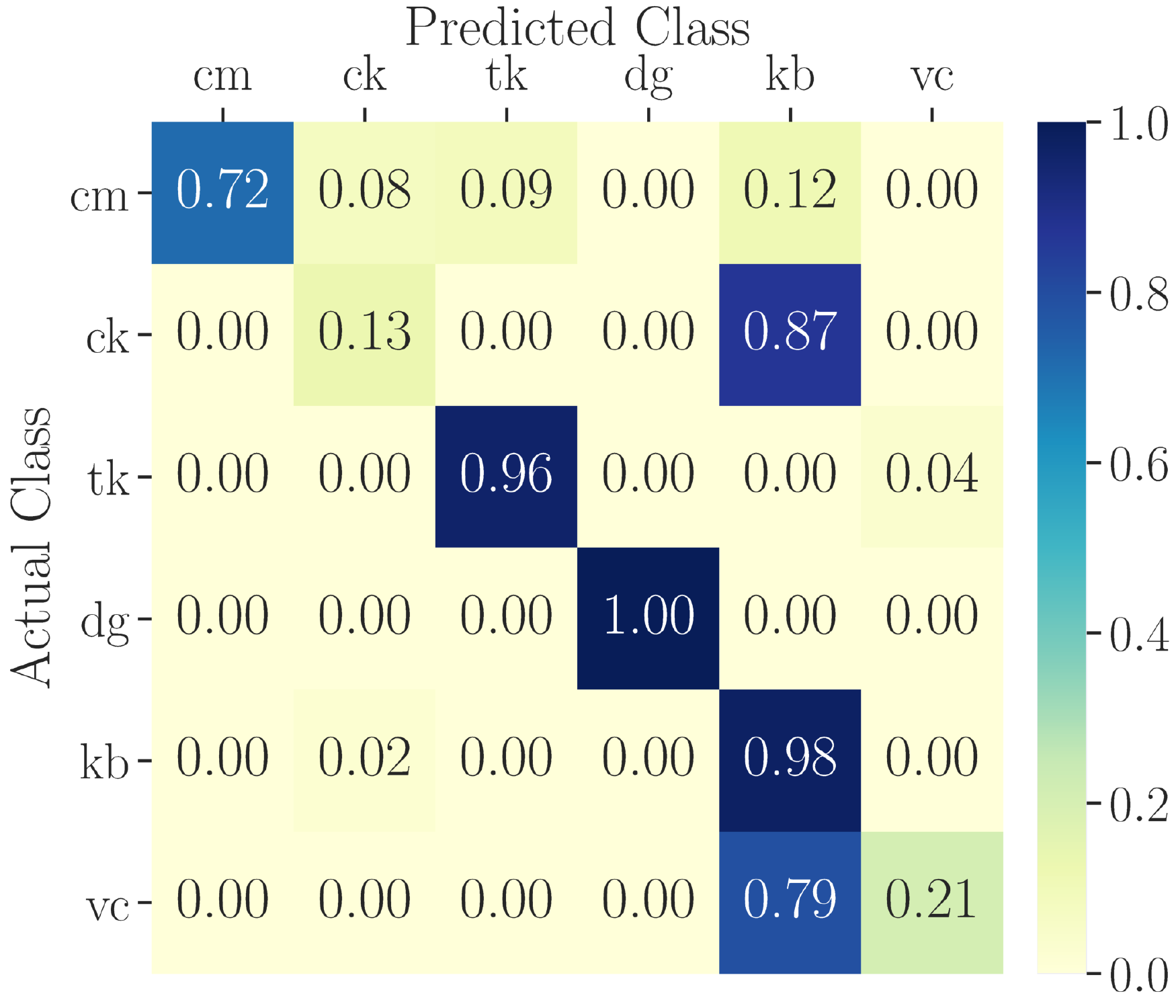}
         \caption{Evalset2win10}
         \label{fig:eval2_10}
     \end{subfigure}
        \caption{Background activity classifier performance with window length = 10. The six classes include Classical Music (cm), Cooking or eating (ck), Talking (tk), Dog Barking (dg), Keyboard (kb),  Vacuum or Cleaning (vc).}
        \label{fig:cofm_bg_win10}
\end{figure*}

We present the confusion matrix of window length 5 and 10 in Fig. \ref{fig:cofm_bg_win10} and Fig. \ref{fig:cofm_bg_win5}.

\begin{figure*}[h]
     \centering
     \captionsetup[subfigure]{justification=centering}
     \begin{subfigure}[b]{0.3\textwidth}
         \centering
         \includegraphics[width=\textwidth]{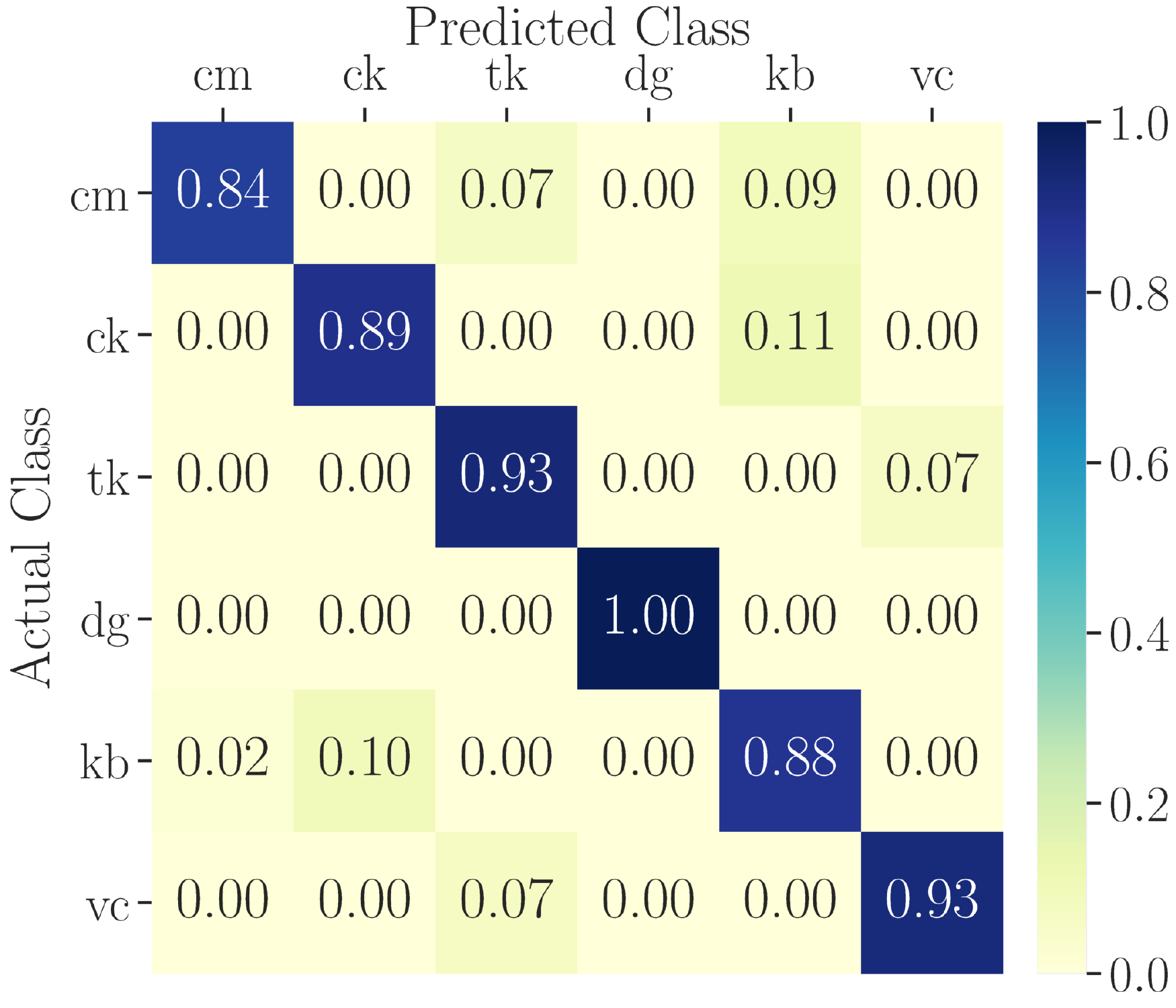}
         \caption{Validation Set}
         \label{fig:valdset_win3}
     \end{subfigure}
     \hfill
     \begin{subfigure}[b]{0.3\textwidth}
         \centering
         \includegraphics[width=\textwidth]{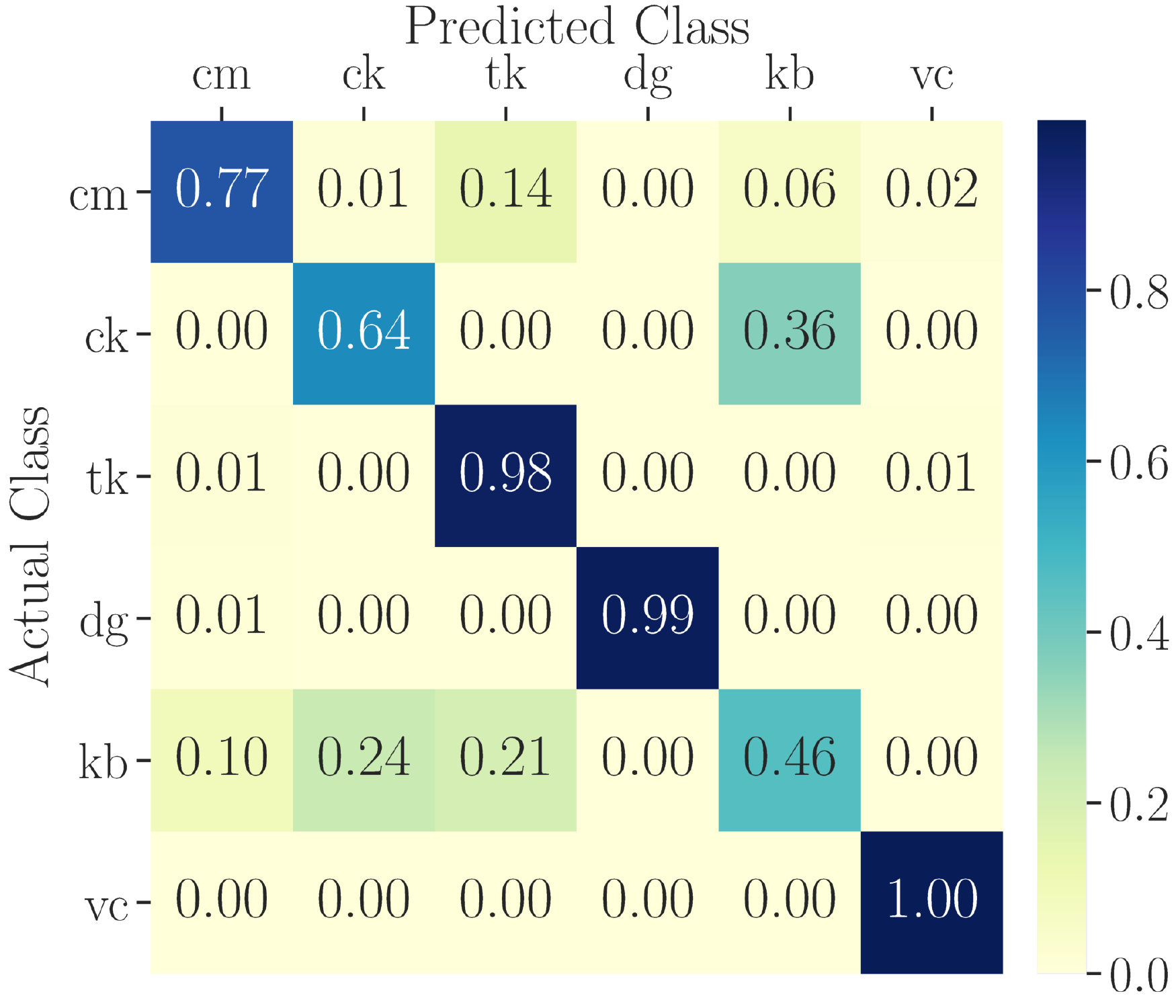}
         \caption{Evalset1}
         \label{fig:eval1_win3}
     \end{subfigure}
     \hfill
     \begin{subfigure}[b]{0.3\textwidth}
         \centering
         \includegraphics[width=\textwidth]{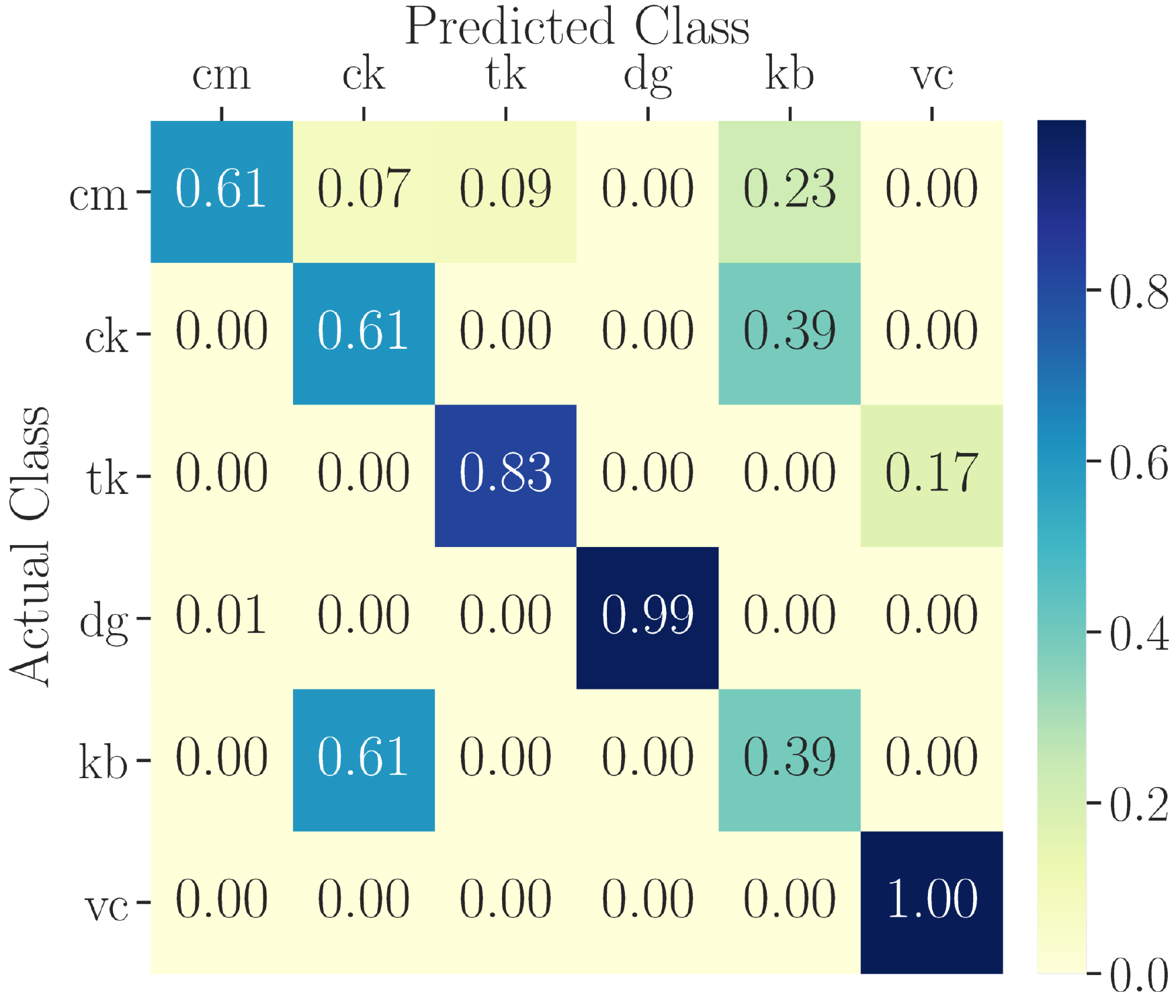}
         \caption{Evalset2}
         \label{fig:eval2_win3}
     \end{subfigure}
        \caption{Background activity classifier performance with window length = 3. The six classes include Classical Music (cm), Cooking or eating (ck), Talking (tk), Dog Barking (dg), Keyboard (kb),  Vacuum or Cleaning (vc).}
        \label{fig:cofm_bg_win3}
\end{figure*}

% \section{Example Webex JSON Array}
% \input{PoPETS2022/jsonexample}
\end{appendix}